\documentclass[twocolumn,showpacs,preprintnumbers,amsmath,amssymb,prl,unsortedaddress,superscriptaddress,longbibliography,letterpaper]{revtex4-1}
\usepackage{graphicx}
\usepackage{verbatim} 
\usepackage{braket}
\usepackage[english]{babel}
\usepackage{xfrac}

\begin{document}


\title{Time dependence of few-body F\"{o}rster interactions among ultracold Rydberg atoms}

\author{Zhimin Cheryl Liu}%
\affiliation{Department of Physics, Bryn Mawr College, Bryn Mawr, PA 19010.}
\affiliation{Department of Physics, University of Colorado, Boulder, CO 80309.}

\author{Nina P. Inman}%
\affiliation{Department of Physics, Bryn Mawr College, Bryn Mawr, PA 19010.}

\author{Thomas J. Carroll}
\affiliation{Department of Physics and Astronomy, Ursinus College, Collegeville, PA 19426.}

\author{Michael W. Noel}%
\affiliation{Department of Physics, Bryn Mawr College, Bryn Mawr, PA 19010.}

\date{\today}

\begin{abstract}
Rubidium Rydberg atoms in either $|m_j|$-sublevel of the $36p_{3/2}$ state can exchange energy via Stark-tuned F\"{o}rster resonances, including two-, three-, and four-body dipole-dipole interactions. Three-body interactions of this type were first reported and categorized by Faoro, \textit{et al.}~[Nat.\ Commun.\ \textbf{6}, 8173 (2015)] and their Borromean nature was confirmed by Tretyakov, \textit{et al.}~[Phys.\ Rev.\ Lett. \textbf{119}, 173402 (2017)]. We report the time dependence of the $N$-body F\"{o}rster resonance $N\times 36p_{3/2,|m_j|=1/2}\rightarrow 36s_{1/2}+37s_{1/2}+(N-2)\times 36p_{3/2,|m_j|=3/2}$, for $N=2,3$, and 4, by measuring the fraction of initially excited atoms that end up in the $37s_{1/2}$ state as a function of time. The essential features of these interactions are captured in an analytical model that includes only the many-body matrix elements and neighboring atom distribution.  A more sophisticated simulation reveals the importance of beyond-nearest-neighbor interactions and of always-resonant interactions.
\end{abstract}


\maketitle

Understanding few-body and many-body interactions is of near universal importance, with relevance to problems in atomic, condensed matter, and nuclear physics. Experiments with ultracold atoms and molecules have significantly advanced that understanding. For example, the spin lattice that forms when polar molecules are confined with an optical lattice may be useful in modeling quantum magnetism and topological insulators~\cite{yan_observation_2013,bohn_cold_2017}. Precise control over the interactions in systems of ultracold atoms has recently been realized in a variety of experiments.  The few-body universal quantum states predicted by Efimov~\cite{efimov_energy_1970} have been observed and studied extensively in ultracold gases~\cite{kraemer_evidence_2006,greene_universal_2017}.  Progress with dipolar quantum gases includes the observation of stable quantum droplets in a dysprosium Bose-Einstein condensate~\cite{ferrier-barbut_observation_2016}, the observation of angular oscillations of quantum droplets, analagous to the behavior of nuclei, induced by the dipole-dipole interaction~\cite{ferrier-barbut_scissors_2018}, and the discovery of a regime with super-solid properties~\cite{tanzi_observation_2019}.  

Dipole-dipole mediated energy exchange in an amorphous ultracold Rydberg gas has been studied extensively over the past two decades~\cite{anderson_resonant_1998,mourachko_many-body_1998}.  Precise lineshape measurements have contributed to our understanding of the importance of many-body and always resonant exchange in this system~\cite{sun_spectral_2008,richards_dipole-dipole_2016,mourachko_controlled_2004,carroll_many-body_2006,yakshina_line_2016}.  Recently, resonant energy transfer between Rydberg atoms and polar molecules has also been observed~\cite{zhelyazkova_electrically_2017,jarisch_state_2018}.

Less attention has been given to the time evolution of resonant energy exchange in this system.  A Ramsey interferency measurement was used to explore dephasing due to always resonant processes \cite{anderson_dephasing_2002}.  Rabi oscillations in the energy exchange between a pair of isolated atoms has been seen \cite{ravets_coherent_2014,ravets_measurement_2015} along with energy exchange between two well-separated macroscopic samples \cite{van_ditzhuijzen_spatially_2008}.  Computational and experimental results that image the time dependence of the energy exchange hint at the possibility of localization in this system \cite{fahey_imaging_2015,bigelow_simulations_2016,robicheaux_effect_2014}.  Using a microwave field to initiate a quantum quench Orioli,~\textit{et al}.\  have explored relaxation of an  ultracold Rydberg gas \cite{orioli_relaxation_2018}.  Further exploration of the time evolution of energy exchange in ultracold Rydberg gases may ultimately shed light on many-body localization and thermalization that complements recent work in other spin systems \cite{kucsko_critical_2018,xu_emulating_2018,smith_many-body_2016}.



Resonant few-body dipole-dipole interactions with Rydberg atoms were discovered and studied only recently. Gurian, \textit{et al}.\ observed a four-body resonant interaction in cesium that lies between, and relies on, a pair of two-body interactions~\cite{gurian_observation_2012}. Faoro, \textit{et al}.\ reported on a simpler three-body process in cesium and developed a theory for a class of similar few-body interactions applicable to many Rydberg atoms~\cite{faoro_borromean_2015}. More recently, Tretyakov, \textit{et al}.\ observed the same type of three-body interaction in rubidium Rydberg atoms while conclusively demonstrating the Borromean nature of the energy exchange~\cite{tretyakov_observation_2017}. Additional work has studied coherence of this interaction and its suitability for use in a quantum gate~\cite{beterov_fast_2018,ryabtsev_coherence_2018}.

In this Letter, we report on the time dependence of the two-, three-, and four-body dipole-dipole interactions in rubidium
\begin{eqnarray}
p + p &\rightarrow& s + s' \label{eq:2body}\\
p + p + p &\rightarrow& s + s' + p'\label{eq:3body}\\
p + p + p + p &\rightarrow& s + s' + p' + p' \label{eq:4body},
\end{eqnarray}
where the state labels have been abbreviated $p=36p_{3/2,|m_j|=1/2}$, $s=37s_{1/2}$, $p'=36p_{3/2,|m_j|=3/2}$, and $s'=36s_{1/2}$. We present a simple model of the time evolution that is able to approximately describe the shape of the time dependence curve, though it neglects always-resonant and beyond-nearest-neighbor interactions. In contrast, a full many-body simulation matches the experiment more closely, revealing the importance of these physical processes.

\begin{figure}
	\centering
	\includegraphics{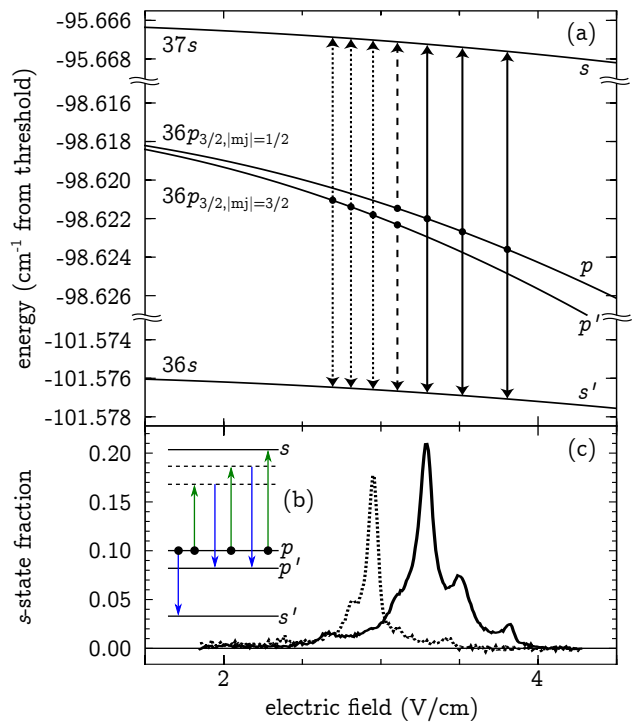}
	\caption{(Color online) (a) Stark map showing the $s$, $p$, $s'$ and $p'$ energy levels as a function of electric field. Solid arrows correspond to F\"{o}rster resonances for which the time dependence was studied, with atoms are initially excited to the $p$ state. Dotted lines are the complementary set of resonances for an initial state of all $p'$ atoms. The dashed line is the location of the two-body $p+p'\rightarrow s + s'$ resonance. (b) Energy level diagram of a possible four-body interaction of Eq.~(\ref{eq:4body}). (c) Experimental $s$-state fraction as a function of electric field for an initial state composed of $p$ atoms (solid line) or $p'$ atoms (dotted line).}	
	\label{fig:starkmap}
\end{figure}

The resonant interactions of Eqs.~(\ref{eq:2body})-(\ref{eq:4body}) are indicated by solid arrows in the Stark map of Fig.~\ref{fig:starkmap}(a). The two-body exchange is resonant at an electric field of 3.29~V/cm. Tuning to higher field introduces an energy defect that is equal to $E_p-E_{p'}$ at 3.52~V/cm. A third $p$ atom can account for the defect via either of the equally detuned two-body exchanges $p + s' \rightarrow s' + p'$ or $p + s \rightarrow s + p'$. Similarly, the energy defect is $2(E_p-E_{p'})$ at 3.80~V/cm which requires a fourth $p$ atom. One example of the many possible four-body interactions is shown in Fig.~\ref{fig:starkmap}(b). While the few-body energy exchange can be perturbatively calculated as a sequence of two-body interactions, it is essentially Borromean in nature and requires all atoms to participate simultaneously~\cite{tretyakov_observation_2017}. More F\"{o}rster resonances following this pattern are possible and are discussed in Ref.~\cite{faoro_borromean_2015}.

In our experiment, about $10^6$ $^{85}$Rb atoms are trapped in a magneto-optical trap (MOT) of diameter $\approx0.5$~mm. The trapping laser at 780~nm cycles atoms between the $5s$ and $5p$ states.  A 776~nm laser drives the  $5p$ to $5d$ transition and a 1265~nm laser excites atoms to the $36p$ state. Simulation suggests a Rydberg density on the order of $10^8$~cm$^{-3}$, corresponding to an average spacing of about 20~$\mu$m. Highly excited atoms then exchange energy through a dipole-dipole interaction and the fraction of atoms in each state is quantified using directed field ionization (DFI)~\cite{gregoric_quantum_2017}.

A set of coaxial cylinders placed on either side of the MOT allow us to apply static and time varying electric fields \cite{gregoric_improving_2018}.  To separate the $p$ and $p'$ states, atoms are excited in an electric field of 4.2~V/cm.  The interaction pulse, which is a square voltage pulse whose length and amplitude can be varied, is then applied to one cylinder.

To determine the strength of this energy exchange, we measure the fraction of atoms that end up in the $s$ state. The time-resolved field ionization signals from the $s$ and $p$ states obtained using standard selective field ionization (SFI) are almost completely overlapping. In addition, the $p$-state signal causes ringing in our detector, which makes quantitative measurement of the $s'$-state fraction difficult. We therefore use DFI to better resolve the $s$ and $p$ states and measure $s$-state fraction~\cite{gregoric_quantum_2017,gregoric_improving_2018,gregoric_perturbed_2019}. Using a genetic algorithm, DFI optimizes a small perturbation that is added to an SFI ramp.  This perturbation directs a fraction of the $s$-state signal along a pathway through the Stark map that ionizes early in time relative to that of the $p$-state signal, allowing us to quantify the fraction of atoms that end up in the $s$ state. During optimization of the DFI perturbation, the delay between excitation and the start of the field ionization ramp is set to zero and no interaction pulse is present.  After optimization we add a fixed delay of 10~$\mu$s between excitation and DFI to provide room for an interaction pulse of varying amplitude and/or length.  During this 10~$\mu$s window, blackbody radiation can drive transitions to neighboring states. This leads to a constant 1.5\% transfer to the $s$ state in the absence of any dipole-dipole interaction, which we subtract from our dipole-dipole fraction measurements.

With the width of the interaction pulse fixed at 9~$\mu$s, we scan the amplitude to tune various interactions into resonance.  We alternate between exciting the $p$ and $p'$ states and average several thousand shots for each amplitude. For a sample excited to the $p$ state, the fraction of atoms that end up in the $s$ state is shown in Fig.~\ref{fig:starkmap}(c) by the solid line.  In this scan we can clearly identify two-, three- and four-body resonant energy exchanges. We also see a feature at the location of the $p+p' \rightarrow s+s'$ resonance, which is marked by the dashed arrows in Fig.~\ref{fig:starkmap}(a).  We associate this with an off-resonant energy exchange.  First the off-resonant $p+p \rightarrow s+s'$ populates the $s+s'$ state.  From this state, the resonant $s+s' \rightarrow p+p'$  can proceed. In contrast to the resonant few-body interactions studied here, this is an inefficient off-resonant multi-step process that seeds the $p+p' \rightarrow s+s'$ exchange.  

Other features in the signal have yet to be understood. In particular, we see a broad tail to the low field side of the two-body exchange along with the small peak near 2.7~V/cm.  The dotted lines in Fig.~\ref{fig:starkmap}(a) and Fig.~\ref{fig:starkmap}(c) show the complementary set of resonances that occur with initial excitation to the $p'$ state. The field axis is calibrated by fitting the locations of the two-body resonances. We check this calibration by measuring the splitting between the $p$ and $p'$ states. These two calibrations agree to within 5\%.

\begin{figure}
	\centering
	\includegraphics{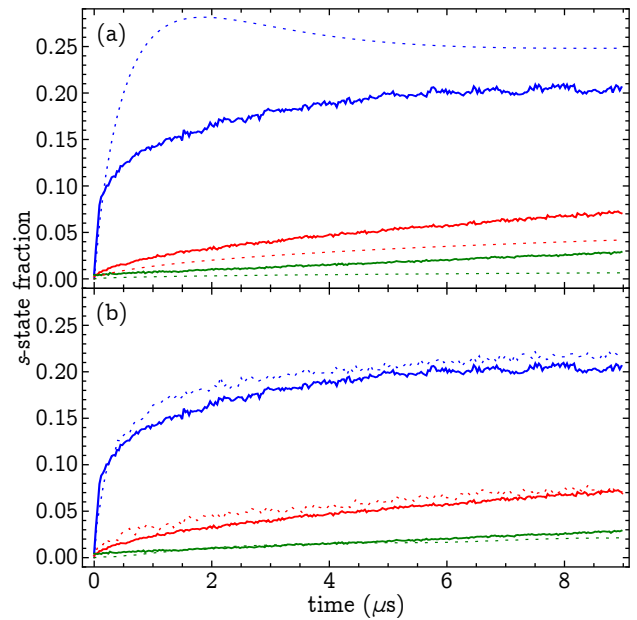}
	\caption{(Color online) Fraction of atoms in the $s$ state as a function of interaction time for (a) the experiment (solid) compared to the simple analytical model (dashed) and (b) the experiment (solid) compared to the simulation (dashed). The two-, \mbox{three-}, and four-body interactions are shown in blue, red, and green respectively. In (a), a density of $9.5\times 10^7$~cm$^{-3}$ was chosen for the simple model to match the initial slope of the two-body interaction. In (b), the simulations were run at a density of $2.4\times 10^8$~cm$^{-3}$. 
	}
	\label{fig:timedep}
\end{figure}

We have also investigated the time dependence of the two-, three-, and four-body interactions.  To collect this data, we scan the time that the interaction field is applied for each of the three resonant fields at 3.29, 3.52, and 3.80~V/cm.  The fraction of atoms in the $s$ state as a function of time is shown by the solid lines in Fig.~\ref{fig:timedep} for each of these three fields. The primary uncertainty in the $s$-state fraction is due to systematic errors in calibration of about 5\%. This is larger than the statistical error since we average over thousands of shots at each time.

For two atoms at fixed separation, the time dependence should be given by Rabi oscillations; in fact, Ravets, \textit{et al}.\ have observed Rabi oscillations between a pair of Rydberg atoms~\cite{ravets_measurement_2015}. A more complicated, but still coherent, oscillation is expected for few-body interactions in a close triplet or quadruplet~\cite{ryabtsev_coherence_2018}. However, we have a different amorphous sample of atoms on each shot of our laser, which averages out the oscillations. The early time behavior of Fig.~\ref{fig:timedep} is dominated by high frequency oscillations among closely spaced atoms, which drive the relatively rapid increase in $s$-state fraction. This is followed by a gradual approach to saturation due to more distant interactions.

We can begin to understand the shape of the time dependence curves for the $N$-body interactions by considering three main factors. First, as $N$ increases, the saturation level of the population transfer decreases so that one expects the $s$-state fraction to eventually plateau at 0.25, 0.1$\overline{6}$, and 0.125 for $N=2$, $N=3$, and $N=4$ respectively. Second, the matrix elements decrease as $N$ increases since each additional two-body step brings in another factor of the detuning. This is, however, somewhat mitigated by the increasing number of paths from initial state to final state as $N$ increases. Finally, in an amorphous sample of atoms the distance between a close pair will be less than the average distance among close triplets or quadruplets. 

We construct a simple analytical model comprised of these three factors. We assume that the $N$-body interaction is dominated by contributions from clusters of $N$ atoms. The $j^{th}$-nearest neighbor probability distribution is given by
\begin{equation}
g(r_{0j}~\vert~r_{0(j-1)})=4\pi r_{0j}^2\rho e^{-\frac{4}{3}\pi\rho(r_{0j}^3-r_{0(j-1)}^3)},
\end{equation}
where $r_{0j}$ is the distance from the central atom to its $j^{th}$-nearest neighbor and $\rho$ is the Rydberg atom density~\cite{edgal_near-neighbor_1983,chandrasekhar_stochastic_1943}. Beyond-nearest-neighbor atoms may be closer to each other than to the central atom. Since the $r^{-3}$ dependence of the dipole-dipole interaction lends much greater weight to closely spaced atoms and $r_{(i\ge 1)j}$ could be significantly less than $r_{0j}$, we average over the distances $r_{(i\ge 1)j}$.

The matrix elements can be calculated perturbatively by summing over all paths from the initial state $\ket{i}$ to the final state $\ket{f}$ with
\begin{eqnarray}
\omega_2&=&\frac{\braket{f|\hat{\sigma}_{fi}|i}}{r_{fi}^3}\label{eq:2me}\\
\omega_3&=&\sum\limits_{j}\frac{\braket{f|\hat{\sigma}_{fj}|j}\braket{j|\hat{\sigma}_{ji}|i}}{\delta_j r_{fj}^3 r_{ji}^3}\label{eq:3me}\\
\omega_4&=&\sum\limits_{j}\sum\limits_{k}\frac{\braket{f|\hat{\sigma}_{fj}|j}\braket{j|\hat{\sigma}_{jk}|k}\braket{k|\hat{\sigma}_{ki}|i}}{\delta_j \delta_k r_{fj}^3 r_{jk}^3 r_{ki}^3}\label{eq:4me},
\end{eqnarray}
where $\ket{j}$ and $\ket{k}$ are intermediate states, $\delta_j$ is the detuning of the $j^{th}$ intermediate state, and $\hat{\sigma}_{ij}$ is the operator that takes the $j^{th}$ state to the $i^{th}$ state. Since the dipole-dipole interaction couples pairs of atoms, each operator $\hat{\sigma}$ represents a product of single-atom operators that take an individual atom from an $s$-state to a $p$-state or vice versa~\cite{bigelow_simulations_2016,younge_mesoscopic_2009}.

For this simple model, we ignore the angular dependence and simply multiply the summands in Eq.~(\ref{eq:3me}) and Eq.~(\ref{eq:4me}) by the total number of possible paths from a given initial to a given final state. We also consider $\delta$ to be fixed for every step, yielding
\begin{equation}
\omega_2 = \frac{\mu\nu}{r_{12}^3}, ~~\omega_3 = \frac{2(\mu\nu)^2}{\delta~r_{12}^3 r_{23}^3 }, ~\mathrm{and~}\omega_4 = \frac{16(\mu\nu)^3}{\delta^2~ r_{12}^3 r_{23}^3 r_{34}^3 },
\end{equation}
where $\mu\approx 700$~$ea_0$ and $\nu \approx 600$~$ea_0$ are the transition dipole moments $p\mathrm{~or~}p' \rightarrow s$ and  $p\mathrm{~or~}p' \rightarrow s'$ respectively. There are significantly more paths from initial to final state if $\pm m_j$ states are included.

The time dependence of the population transfer to the $s$ state among a cluster of $N$ atoms should be oscillatory with frequencies similar to $\omega_N$ and an amplitude determined by the saturation level. Since we average the time dependence over atomic separations, the particular form of the oscillation is not important. We use $(2N)^{-1} \sin^2(\omega_N t)$, where $(2N)^{-1}$ gives the saturation level. The population fractions, $P_N(t)$, transferred to the $s$ state are 
\begin{widetext}
\begin{eqnarray}
P_2(t) &=& 2\pi\rho\int_0^\infty e^{-\frac{4}{3}\pi\rho r^3} \sin^2\Biggl[\frac{\mu\nu}{r^3}t\Biggr]dr,\label{eq:2b}\\
P_3(t) &=& \frac{16\pi\rho^2}{3} \int_0^R r_{02}^2 e^{-\frac{4}{3}\pi\rho r_{02}^3} \int_0^{r_{02}} r_{01}^2\int_0^\pi \sin^2\Biggl[\frac{(\mu\nu)^2 t}{\delta~r_{01}^3(r_{01}^2+r_{02}^2-2r_{01}r_{02}\cos\theta_{12})^{3/2}}\Biggr]d\theta_{12} dr_{01} dr_{02},\label{eq:3b}\\
P_4(t) &=& 16\pi\rho^3\int_0^R r_{03}^2 e^{-\frac{4}{3}\pi\rho r_{03}^3} \int_0^{r_{03}}r_{02}^2 \int_0^{r_{02}}r_{01}^2 \int_0^\pi \int_0^\pi \nonumber\\ 
&~&~~~~\sin^2\Biggl[\frac{(\mu\nu)^3 t}{\delta^2~r_{01}^3(r_{01}^2+r_{02}^2-2r_{01}r_{02}\cos\theta_{12})^{3/2}(r_{02}^2+r_{03}^2-2r_{02}r_{03}\cos\theta_{23})^{3/2}}\Biggr]d\theta_{12}d\theta_{23} dr_{01} dr_{02} dr_{03},\label{eq:4b}
\end{eqnarray}
\end{widetext}
where $\theta_{ij}$ is the angle between $\vec{r}_{0i}$ and $\vec{r}_{0j}$ and the numerical pre-factor includes the saturation level. The two-body result of Eq.~(\ref{eq:2b}) can be integrated analytically~\cite{yakshina_line_2016,cournol_accurate_2018}. The three- and four-body results in Eqs.~(\ref{eq:3b})-(\ref{eq:4b}) must be integrated numerically. The integration is terminated at a radius $R$ large enough that it converges.

The time dependence of Eqs.~(\ref{eq:2b})-(\ref{eq:4b}) is shown by dashed lines in Fig.~\ref{fig:timedep}(a). The density was chosen to match the initial slope of the model's two-body time dependence to the experimental data, yielding $\rho=1.0(0.3)\times 10^8$~cm$^{-3}$. This model captures the essential features of the experimental curves. However, in attempting to match the two-body data, it significantly under-predicts the three- and four-body population transfer. This simple model ignores the fact that the experimental three- and four-body time-dependence each include a contribution from the shoulder of the two-body interaction. This could account for some, but not all, of the disagreement.

Our simple model neglects always-resonant interactions and beyond-nearest-neighbor interactions, which have been shown to play a role in the dipole-dipole energy exchange in a three-Rydberg-atom chain~\cite{barredo_coherent_2015}. To address these deficiencies, we have also simulated our system by constructing the Hamiltonian matrix using Eqs.~(\ref{eq:2me})-(\ref{eq:4me}) while averaging over the angular dependence~\cite{carroll_angular_2004,ravets_measurement_2015}, under the assumption that the Rydberg atoms remain frozen in place for the duration of the experiment. We randomly place 40 atoms in a spherical volume of radius 34~$\mu$m, while using the blockade radius to estimate a minimum distance between atoms. We calculate the time evolution by solving the Schr\"{o}dinger equation on a supercomputer for each atom and its closest 8 neighbors, resulting in a Hamiltonian matrix of rank 48620. The results are averaged over all 40 atoms and the process is repeated so that a few hundred random instances are averaged. The simulated results, which agree well with the data, are shown by the dashed lines in Fig.~\ref{fig:timedep}(b).

Our simulations suggest that our Rydberg density is $2.4(0.1)\times 10^8$~cm$^{-3}$, where the statistical uncertainty was calculated from multiple simulation runs at a range of densities. By averaging over many random arrangements of atoms at this density, we calculate average interaction strengths of $\omega_2=162\pm 6$~kHz, $\omega_3=1.6\pm 0.1$~kHz, and $\omega_4=0.7\pm 0.1$~kHz for the two-, three-, and four-body interactions, respectively. Since each of the resonant interactions  scale differently with density, one could potentially fit simulations to the data to measure the Rydberg density of the sample. 

Our experiment is a quantum quench~\cite{polkovnikov_colloquium_2011,abanin_recent_2017}. We initially excite atoms to a many-body eigenstate of the Stark Hamiltonian at an electric field where they are non-interacting and then switch to the dipole-dipole interaction Hamiltonian by changing the field. The subsequent time evolution can be broadly divided into two outcomes. The system can thermalize, eventually achieving an equilibrium state that can be specified with a small number of parameters. Or, in the case of many-body localization, the system fails to thermalize and retains a memory of its initial state~\cite{nandkishore_many-body_2015,abanin_recent_2017,abanin_colloquium_2019}. Quenches have been used extensively to study both many-body localization and thermalization~\cite{orioli_relaxation_2018,abanin_colloquium_2019,lukin_probing_2019,sous_possible_2018,sous_many-body_2019}.

Since long-range dipole-dipole interactions should lead to efficient energy transport, one expects our system to thermalize. Indeed, for the two-body interaction this is what we seem to observe as the $s$-state fraction saturates near the expected level of 0.25. However, for the three-body and especially the four-body interactions, the saturation level is significantly lower than expected. This could indicate that the system fails to thermalize. In fact, Nandkishore and Sondhi have recently shown that many-body localization could be possible even in systems with long-range interactions~\cite{nandkishore_many-body_2017}. 

To further investigate, we have extended our simulations of the four-body interaction to longer times and higher densities. The results show that the $s$-state fraction does not saturate at the expected value of 0.125. T\'{a}vora, \textit{et al}.~\cite{tavora_inevitable_2016,tavora_power-law_2017}, suggest using the survival probability of the initial state as a criterion for numerically predicting thermalization. A rapidly decaying survival probability is a sign of thermalization as the memory of the initial state becomes inaccessible due to the spread of entanglement throughout the system. Our preliminary numerical analysis shows that the initial state survival probability in the four-body case does, indeed, decay significantly more slowly than for the two-body interaction. 

We have presented experimental data showing the time dependence of few-body interactions in an amorphous ultracold sample of Rydberg atoms. While the matrix elements for the three- and four-body interactions are reduced because of the detuning, these interactions are stronger than one might expect because of the many paths from initial state to final state. The densities extracted from our simple model and our simulation differ by about a factor of two or three, revealing the importance of always-resonant and beyond-nearest-neighbor interactions. Finally, the population transfer saturation levels suggest that the system may not thermalize as expected. Since the four-body resonance is relatively well-separated from the two- and three-body peaks, it could prove useful for future experiments studying thermalization and localization.

\begin{acknowledgments}
This work was supported by the National Science Foundation under Grants No. 1607335 and No. 1607377.
\end{acknowledgments}

\bibliography{pp2ss-time}

\begin{thebibliography}{52}%
\makeatletter
\providecommand \@ifxundefined [1]{%
 \@ifx{#1\undefined}
}%
\providecommand \@ifnum [1]{%
 \ifnum #1\expandafter \@firstoftwo
 \else \expandafter \@secondoftwo
 \fi
}%
\providecommand \@ifx [1]{%
 \ifx #1\expandafter \@firstoftwo
 \else \expandafter \@secondoftwo
 \fi
}%
\providecommand \natexlab [1]{#1}%
\providecommand \enquote  [1]{``#1''}%
\providecommand \bibnamefont  [1]{#1}%
\providecommand \bibfnamefont [1]{#1}%
\providecommand \citenamefont [1]{#1}%
\providecommand \href@noop [0]{\@secondoftwo}%
\providecommand \href [0]{\begingroup \@sanitize@url \@href}%
\providecommand \@href[1]{\@@startlink{#1}\@@href}%
\providecommand \@@href[1]{\endgroup#1\@@endlink}%
\providecommand \@sanitize@url [0]{\catcode `\\12\catcode `\$12\catcode
  `\&12\catcode `\#12\catcode `\^12\catcode `\_12\catcode `\%12\relax}%
\providecommand \@@startlink[1]{}%
\providecommand \@@endlink[0]{}%
\providecommand \url  [0]{\begingroup\@sanitize@url \@url }%
\providecommand \@url [1]{\endgroup\@href {#1}{\urlprefix }}%
\providecommand \urlprefix  [0]{URL }%
\providecommand \Eprint [0]{\href }%
\providecommand \doibase [0]{http://dx.doi.org/}%
\providecommand \selectlanguage [0]{\@gobble}%
\providecommand \bibinfo  [0]{\@secondoftwo}%
\providecommand \bibfield  [0]{\@secondoftwo}%
\providecommand \translation [1]{[#1]}%
\providecommand \BibitemOpen [0]{}%
\providecommand \bibitemStop [0]{}%
\providecommand \bibitemNoStop [0]{.\EOS\space}%
\providecommand \EOS [0]{\spacefactor3000\relax}%
\providecommand \BibitemShut  [1]{\csname bibitem#1\endcsname}%
\let\auto@bib@innerbib\@empty
\bibitem [{\citenamefont {Yan}\ \emph {et~al.}(2013)\citenamefont {Yan},
  \citenamefont {Moses}, \citenamefont {Gadway}, \citenamefont {Covey},
  \citenamefont {Hazzard}, \citenamefont {Rey}, \citenamefont {Jin},\ and\
  \citenamefont {Ye}}]{yan_observation_2013}%
  \BibitemOpen
  \bibfield  {author} {\bibinfo {author} {\bibfnamefont {Bo}~\bibnamefont
  {Yan}}, \bibinfo {author} {\bibfnamefont {Steven~A.}\ \bibnamefont {Moses}},
  \bibinfo {author} {\bibfnamefont {Bryce}\ \bibnamefont {Gadway}}, \bibinfo
  {author} {\bibfnamefont {Jacob~P.}\ \bibnamefont {Covey}}, \bibinfo {author}
  {\bibfnamefont {Kaden R.~A.}\ \bibnamefont {Hazzard}}, \bibinfo {author}
  {\bibfnamefont {Ana~Maria}\ \bibnamefont {Rey}}, \bibinfo {author}
  {\bibfnamefont {Deborah~S.}\ \bibnamefont {Jin}}, \ and\ \bibinfo {author}
  {\bibfnamefont {Jun}\ \bibnamefont {Ye}},\ }\bibfield  {title} {\enquote
  {\bibinfo {title} {Observation of dipolar spin-exchange interactions with
  lattice-confined polar molecules},}\ }\href {\doibase 10.1038/nature12483}
  {\bibfield  {journal} {\bibinfo  {journal} {Nature}\ }\textbf {\bibinfo
  {volume} {501}},\ \bibinfo {pages} {521--525} (\bibinfo {year}
  {2013})}\BibitemShut {NoStop}%
\bibitem [{\citenamefont {Bohn}\ \emph {et~al.}(2017)\citenamefont {Bohn},
  \citenamefont {Rey},\ and\ \citenamefont {Ye}}]{bohn_cold_2017}%
  \BibitemOpen
  \bibfield  {author} {\bibinfo {author} {\bibfnamefont {John~L.}\ \bibnamefont
  {Bohn}}, \bibinfo {author} {\bibfnamefont {Ana~Maria}\ \bibnamefont {Rey}}, \
  and\ \bibinfo {author} {\bibfnamefont {Jun}\ \bibnamefont {Ye}},\ }\bibfield
  {title} {\enquote {\bibinfo {title} {Cold molecules: {{Progress}} in quantum
  engineering of chemistry and quantum matter},}\ }\href {\doibase
  10.1126/science.aam6299} {\bibfield  {journal} {\bibinfo  {journal}
  {Science}\ }\textbf {\bibinfo {volume} {357}},\ \bibinfo {pages} {1002--1010}
  (\bibinfo {year} {2017})}\BibitemShut {NoStop}%
\bibitem [{\citenamefont {Efimov}(1970)}]{efimov_energy_1970}%
  \BibitemOpen
  \bibfield  {author} {\bibinfo {author} {\bibfnamefont {V.}~\bibnamefont
  {Efimov}},\ }\bibfield  {title} {\enquote {\bibinfo {title} {Energy levels
  arising from resonant two-body forces in a three-body system},}\ }\href
  {\doibase 10.1016/0370-2693(70)90349-7} {\bibfield  {journal} {\bibinfo
  {journal} {Physics Letters B}\ }\textbf {\bibinfo {volume} {33}},\ \bibinfo
  {pages} {563--564} (\bibinfo {year} {1970})}\BibitemShut {NoStop}%
\bibitem [{\citenamefont {Kraemer}\ \emph {et~al.}(2006)\citenamefont
  {Kraemer}, \citenamefont {Mark}, \citenamefont {Waldburger}, \citenamefont
  {Danzl}, \citenamefont {Chin}, \citenamefont {Engeser}, \citenamefont
  {Lange}, \citenamefont {Pilch}, \citenamefont {Jaakkola}, \citenamefont
  {N{\"a}gerl},\ and\ \citenamefont {Grimm}}]{kraemer_evidence_2006}%
  \BibitemOpen
  \bibfield  {author} {\bibinfo {author} {\bibfnamefont {T.}~\bibnamefont
  {Kraemer}}, \bibinfo {author} {\bibfnamefont {M.}~\bibnamefont {Mark}},
  \bibinfo {author} {\bibfnamefont {P.}~\bibnamefont {Waldburger}}, \bibinfo
  {author} {\bibfnamefont {J.~G.}\ \bibnamefont {Danzl}}, \bibinfo {author}
  {\bibfnamefont {C.}~\bibnamefont {Chin}}, \bibinfo {author} {\bibfnamefont
  {B.}~\bibnamefont {Engeser}}, \bibinfo {author} {\bibfnamefont {A.~D.}\
  \bibnamefont {Lange}}, \bibinfo {author} {\bibfnamefont {K.}~\bibnamefont
  {Pilch}}, \bibinfo {author} {\bibfnamefont {A.}~\bibnamefont {Jaakkola}},
  \bibinfo {author} {\bibfnamefont {H.-C.}\ \bibnamefont {N{\"a}gerl}}, \ and\
  \bibinfo {author} {\bibfnamefont {R.}~\bibnamefont {Grimm}},\ }\bibfield
  {title} {\enquote {\bibinfo {title} {Evidence for {{Efimov}} quantum states
  in an ultracold gas of caesium atoms},}\ }\href {\doibase
  10.1038/nature04626} {\bibfield  {journal} {\bibinfo  {journal} {Nature}\
  }\textbf {\bibinfo {volume} {440}},\ \bibinfo {pages} {315--318} (\bibinfo
  {year} {2006})}\BibitemShut {NoStop}%
\bibitem [{\citenamefont {Greene}\ \emph {et~al.}(2017)\citenamefont {Greene},
  \citenamefont {Giannakeas},\ and\ \citenamefont
  {{P{\'e}rez-R{\'i}os}}}]{greene_universal_2017}%
  \BibitemOpen
  \bibfield  {author} {\bibinfo {author} {\bibfnamefont {Chris~H.}\
  \bibnamefont {Greene}}, \bibinfo {author} {\bibfnamefont {P.}~\bibnamefont
  {Giannakeas}}, \ and\ \bibinfo {author} {\bibfnamefont {J.}~\bibnamefont
  {{P{\'e}rez-R{\'i}os}}},\ }\bibfield  {title} {\enquote {\bibinfo {title}
  {Universal few-body physics and cluster formation},}\ }\href {\doibase
  10.1103/RevModPhys.89.035006} {\bibfield  {journal} {\bibinfo  {journal}
  {Rev. Mod. Phys.}\ }\textbf {\bibinfo {volume} {89}},\ \bibinfo {pages}
  {035006} (\bibinfo {year} {2017})}\BibitemShut {NoStop}%
\bibitem [{\citenamefont {{Ferrier-Barbut}}\ \emph {et~al.}(2016)\citenamefont
  {{Ferrier-Barbut}}, \citenamefont {Kadau}, \citenamefont {Schmitt},
  \citenamefont {Wenzel},\ and\ \citenamefont
  {Pfau}}]{ferrier-barbut_observation_2016}%
  \BibitemOpen
  \bibfield  {author} {\bibinfo {author} {\bibfnamefont {Igor}\ \bibnamefont
  {{Ferrier-Barbut}}}, \bibinfo {author} {\bibfnamefont {Holger}\ \bibnamefont
  {Kadau}}, \bibinfo {author} {\bibfnamefont {Matthias}\ \bibnamefont
  {Schmitt}}, \bibinfo {author} {\bibfnamefont {Matthias}\ \bibnamefont
  {Wenzel}}, \ and\ \bibinfo {author} {\bibfnamefont {Tilman}\ \bibnamefont
  {Pfau}},\ }\bibfield  {title} {\enquote {\bibinfo {title} {Observation of
  {{Quantum Droplets}} in a {{Strongly Dipolar Bose Gas}}},}\ }\href {\doibase
  10.1103/PhysRevLett.116.215301} {\bibfield  {journal} {\bibinfo  {journal}
  {Phys. Rev. Lett.}\ }\textbf {\bibinfo {volume} {116}},\ \bibinfo {pages}
  {215301} (\bibinfo {year} {2016})}\BibitemShut {NoStop}%
\bibitem [{\citenamefont {{Ferrier-Barbut}}\ \emph {et~al.}(2018)\citenamefont
  {{Ferrier-Barbut}}, \citenamefont {Wenzel}, \citenamefont {B{\"o}ttcher},
  \citenamefont {Langen}, \citenamefont {Isoard}, \citenamefont {Stringari},\
  and\ \citenamefont {Pfau}}]{ferrier-barbut_scissors_2018}%
  \BibitemOpen
  \bibfield  {author} {\bibinfo {author} {\bibfnamefont {Igor}\ \bibnamefont
  {{Ferrier-Barbut}}}, \bibinfo {author} {\bibfnamefont {Matthias}\
  \bibnamefont {Wenzel}}, \bibinfo {author} {\bibfnamefont {Fabian}\
  \bibnamefont {B{\"o}ttcher}}, \bibinfo {author} {\bibfnamefont {Tim}\
  \bibnamefont {Langen}}, \bibinfo {author} {\bibfnamefont {Mathieu}\
  \bibnamefont {Isoard}}, \bibinfo {author} {\bibfnamefont {Sandro}\
  \bibnamefont {Stringari}}, \ and\ \bibinfo {author} {\bibfnamefont {Tilman}\
  \bibnamefont {Pfau}},\ }\bibfield  {title} {\enquote {\bibinfo {title}
  {Scissors {{Mode}} of {{Dipolar Quantum Droplets}} of {{Dysprosium
  Atoms}}},}\ }\href {\doibase 10.1103/PhysRevLett.120.160402} {\bibfield
  {journal} {\bibinfo  {journal} {Phys. Rev. Lett.}\ }\textbf {\bibinfo
  {volume} {120}},\ \bibinfo {pages} {160402} (\bibinfo {year}
  {2018})}\BibitemShut {NoStop}%
\bibitem [{\citenamefont {Tanzi}\ \emph {et~al.}(2019)\citenamefont {Tanzi},
  \citenamefont {Lucioni}, \citenamefont {Fam{\`a}}, \citenamefont {Catani},
  \citenamefont {Fioretti}, \citenamefont {Gabbanini}, \citenamefont {Bisset},
  \citenamefont {Santos},\ and\ \citenamefont
  {Modugno}}]{tanzi_observation_2019}%
  \BibitemOpen
  \bibfield  {author} {\bibinfo {author} {\bibfnamefont {L.}~\bibnamefont
  {Tanzi}}, \bibinfo {author} {\bibfnamefont {E.}~\bibnamefont {Lucioni}},
  \bibinfo {author} {\bibfnamefont {F.}~\bibnamefont {Fam{\`a}}}, \bibinfo
  {author} {\bibfnamefont {J.}~\bibnamefont {Catani}}, \bibinfo {author}
  {\bibfnamefont {A.}~\bibnamefont {Fioretti}}, \bibinfo {author}
  {\bibfnamefont {C.}~\bibnamefont {Gabbanini}}, \bibinfo {author}
  {\bibfnamefont {R.~N.}\ \bibnamefont {Bisset}}, \bibinfo {author}
  {\bibfnamefont {L.}~\bibnamefont {Santos}}, \ and\ \bibinfo {author}
  {\bibfnamefont {G.}~\bibnamefont {Modugno}},\ }\bibfield  {title} {\enquote
  {\bibinfo {title} {Observation of a {{Dipolar Quantum Gas}} with {{Metastable
  Supersolid Properties}}},}\ }\href {\doibase 10.1103/PhysRevLett.122.130405}
  {\bibfield  {journal} {\bibinfo  {journal} {Phys. Rev. Lett.}\ }\textbf
  {\bibinfo {volume} {122}},\ \bibinfo {pages} {130405} (\bibinfo {year}
  {2019})}\BibitemShut {NoStop}%
\bibitem [{\citenamefont {Anderson}\ \emph {et~al.}(1998)\citenamefont
  {Anderson}, \citenamefont {Veale},\ and\ \citenamefont
  {Gallagher}}]{anderson_resonant_1998}%
  \BibitemOpen
  \bibfield  {author} {\bibinfo {author} {\bibfnamefont {W.~R.}\ \bibnamefont
  {Anderson}}, \bibinfo {author} {\bibfnamefont {J.~R.}\ \bibnamefont {Veale}},
  \ and\ \bibinfo {author} {\bibfnamefont {T.~F.}\ \bibnamefont {Gallagher}},\
  }\bibfield  {title} {\enquote {\bibinfo {title} {Resonant {{Dipole}}-{{Dipole
  Energy Transfer}} in a {{Nearly Frozen Rydberg Gas}}},}\ }\href {\doibase
  10.1103/PhysRevLett.80.249} {\bibfield  {journal} {\bibinfo  {journal} {Phys.
  Rev. Lett.}\ }\textbf {\bibinfo {volume} {80}},\ \bibinfo {pages} {249--252}
  (\bibinfo {year} {1998})}\BibitemShut {NoStop}%
\bibitem [{\citenamefont {Mourachko}\ \emph {et~al.}(1998)\citenamefont
  {Mourachko}, \citenamefont {Comparat}, \citenamefont {{de Tomasi}},
  \citenamefont {Fioretti}, \citenamefont {Nosbaum}, \citenamefont {Akulin},\
  and\ \citenamefont {Pillet}}]{mourachko_many-body_1998}%
  \BibitemOpen
  \bibfield  {author} {\bibinfo {author} {\bibfnamefont {I.}~\bibnamefont
  {Mourachko}}, \bibinfo {author} {\bibfnamefont {D.}~\bibnamefont {Comparat}},
  \bibinfo {author} {\bibfnamefont {F.}~\bibnamefont {{de Tomasi}}}, \bibinfo
  {author} {\bibfnamefont {A.}~\bibnamefont {Fioretti}}, \bibinfo {author}
  {\bibfnamefont {P.}~\bibnamefont {Nosbaum}}, \bibinfo {author} {\bibfnamefont
  {V.~M.}\ \bibnamefont {Akulin}}, \ and\ \bibinfo {author} {\bibfnamefont
  {P.}~\bibnamefont {Pillet}},\ }\bibfield  {title} {\enquote {\bibinfo {title}
  {Many-{{Body Effects}} in a {{Frozen Rydberg Gas}}},}\ }\href {\doibase
  10.1103/PhysRevLett.80.253} {\bibfield  {journal} {\bibinfo  {journal} {Phys.
  Rev. Lett.}\ }\textbf {\bibinfo {volume} {80}},\ \bibinfo {pages} {253--256}
  (\bibinfo {year} {1998})}\BibitemShut {NoStop}%
\bibitem [{\citenamefont {Sun}\ and\ \citenamefont
  {Robicheaux}(2008)}]{sun_spectral_2008}%
  \BibitemOpen
  \bibfield  {author} {\bibinfo {author} {\bibfnamefont {B.}~\bibnamefont
  {Sun}}\ and\ \bibinfo {author} {\bibfnamefont {F.}~\bibnamefont
  {Robicheaux}},\ }\bibfield  {title} {\enquote {\bibinfo {title} {Spectral
  linewidth broadening from pair fluctuations in a frozen {{Rydberg}} gas},}\
  }\href {\doibase 10.1103/PhysRevA.78.040701} {\bibfield  {journal} {\bibinfo
  {journal} {Phys. Rev. A}\ }\textbf {\bibinfo {volume} {78}},\ \bibinfo
  {pages} {040701(R)} (\bibinfo {year} {2008})}\BibitemShut {NoStop}%
\bibitem [{\citenamefont {Richards}\ and\ \citenamefont
  {Jones}(2016)}]{richards_dipole-dipole_2016}%
  \BibitemOpen
  \bibfield  {author} {\bibinfo {author} {\bibfnamefont {B.~G.}\ \bibnamefont
  {Richards}}\ and\ \bibinfo {author} {\bibfnamefont {R.~R.}\ \bibnamefont
  {Jones}},\ }\bibfield  {title} {\enquote {\bibinfo {title} {Dipole-dipole
  resonance line shapes in a cold {{Rydberg}} gas},}\ }\href {\doibase
  10.1103/PhysRevA.93.042505} {\bibfield  {journal} {\bibinfo  {journal} {Phys.
  Rev. A}\ }\textbf {\bibinfo {volume} {93}},\ \bibinfo {pages} {042505}
  (\bibinfo {year} {2016})}\BibitemShut {NoStop}%
\bibitem [{\citenamefont {Mourachko}\ \emph {et~al.}(2004)\citenamefont
  {Mourachko}, \citenamefont {Li},\ and\ \citenamefont
  {Gallagher}}]{mourachko_controlled_2004}%
  \BibitemOpen
  \bibfield  {author} {\bibinfo {author} {\bibfnamefont {I.}~\bibnamefont
  {Mourachko}}, \bibinfo {author} {\bibfnamefont {Wenhui}\ \bibnamefont {Li}},
  \ and\ \bibinfo {author} {\bibfnamefont {T.~F.}\ \bibnamefont {Gallagher}},\
  }\bibfield  {title} {\enquote {\bibinfo {title} {Controlled many-body
  interactions in a frozen {{Rydberg}} gas},}\ }\href {\doibase
  10.1103/PhysRevA.70.031401} {\bibfield  {journal} {\bibinfo  {journal} {Phys.
  Rev. A}\ }\textbf {\bibinfo {volume} {70}},\ \bibinfo {pages} {031401(R)}
  (\bibinfo {year} {2004})}\BibitemShut {NoStop}%
\bibitem [{\citenamefont {Carroll}\ \emph {et~al.}(2006)\citenamefont
  {Carroll}, \citenamefont {Sunder},\ and\ \citenamefont
  {Noel}}]{carroll_many-body_2006}%
  \BibitemOpen
  \bibfield  {author} {\bibinfo {author} {\bibfnamefont {Thomas~J.}\
  \bibnamefont {Carroll}}, \bibinfo {author} {\bibfnamefont {Shubha}\
  \bibnamefont {Sunder}}, \ and\ \bibinfo {author} {\bibfnamefont {Michael~W.}\
  \bibnamefont {Noel}},\ }\bibfield  {title} {\enquote {\bibinfo {title}
  {Many-body interactions in a sample of ultracold {{Rydberg}} atoms with
  varying dimensions and densities},}\ }\href {\doibase
  10.1103/PhysRevA.73.032725} {\bibfield  {journal} {\bibinfo  {journal} {Phys.
  Rev. A}\ }\textbf {\bibinfo {volume} {73}},\ \bibinfo {pages} {032725}
  (\bibinfo {year} {2006})}\BibitemShut {NoStop}%
\bibitem [{\citenamefont {Yakshina}\ \emph {et~al.}(2016)\citenamefont
  {Yakshina}, \citenamefont {Tretyakov}, \citenamefont {Beterov}, \citenamefont
  {Entin}, \citenamefont {Andreeva}, \citenamefont {Cinins}, \citenamefont
  {Markovski}, \citenamefont {Iftikhar}, \citenamefont {Ekers},\ and\
  \citenamefont {Ryabtsev}}]{yakshina_line_2016}%
  \BibitemOpen
  \bibfield  {author} {\bibinfo {author} {\bibfnamefont {E.~A.}\ \bibnamefont
  {Yakshina}}, \bibinfo {author} {\bibfnamefont {D.~B.}\ \bibnamefont
  {Tretyakov}}, \bibinfo {author} {\bibfnamefont {I.~I.}\ \bibnamefont
  {Beterov}}, \bibinfo {author} {\bibfnamefont {V.~M.}\ \bibnamefont {Entin}},
  \bibinfo {author} {\bibfnamefont {C.}~\bibnamefont {Andreeva}}, \bibinfo
  {author} {\bibfnamefont {A.}~\bibnamefont {Cinins}}, \bibinfo {author}
  {\bibfnamefont {A.}~\bibnamefont {Markovski}}, \bibinfo {author}
  {\bibfnamefont {Z.}~\bibnamefont {Iftikhar}}, \bibinfo {author}
  {\bibfnamefont {A.}~\bibnamefont {Ekers}}, \ and\ \bibinfo {author}
  {\bibfnamefont {I.~I.}\ \bibnamefont {Ryabtsev}},\ }\bibfield  {title}
  {\enquote {\bibinfo {title} {Line shapes and time dynamics of the {{Forster}}
  resonances between two {{Rydberg}} atoms in a time-varying electric field},}\
  }\href {\doibase 10.1103/PhysRevA.94.043417} {\bibfield  {journal} {\bibinfo
  {journal} {Phys. Rev. A}\ }\textbf {\bibinfo {volume} {94}},\ \bibinfo
  {pages} {043417} (\bibinfo {year} {2016})}\BibitemShut {NoStop}%
\bibitem [{\citenamefont {Zhelyazkova}\ and\ \citenamefont
  {Hogan}(2017)}]{zhelyazkova_electrically_2017}%
  \BibitemOpen
  \bibfield  {author} {\bibinfo {author} {\bibfnamefont {V.}~\bibnamefont
  {Zhelyazkova}}\ and\ \bibinfo {author} {\bibfnamefont {S.~D.}\ \bibnamefont
  {Hogan}},\ }\bibfield  {title} {\enquote {\bibinfo {title} {Electrically
  tuned {{F{\"o}rster}} resonances in collisions of {${\mathrm{NH}}_{3}$} with
  {{Rydberg He}} atoms},}\ }\href {\doibase 10.1103/PhysRevA.95.042710}
  {\bibfield  {journal} {\bibinfo  {journal} {Phys. Rev. A}\ }\textbf {\bibinfo
  {volume} {95}},\ \bibinfo {pages} {042710} (\bibinfo {year}
  {2017})}\BibitemShut {NoStop}%
\bibitem [{\citenamefont {Jarisch}\ and\ \citenamefont
  {Zeppenfeld}(2018)}]{jarisch_state_2018}%
  \BibitemOpen
  \bibfield  {author} {\bibinfo {author} {\bibfnamefont {F.}~\bibnamefont
  {Jarisch}}\ and\ \bibinfo {author} {\bibfnamefont {M.}~\bibnamefont
  {Zeppenfeld}},\ }\bibfield  {title} {\enquote {\bibinfo {title} {State
  resolved investigation of {{F{\"o}rster}} resonant energy transfer in
  collisions between polar molecules and {{Rydberg}} atoms},}\ }\href {\doibase
  10.1088/1367-2630/aaf02e} {\bibfield  {journal} {\bibinfo  {journal} {New J.
  Phys.}\ }\textbf {\bibinfo {volume} {20}},\ \bibinfo {pages} {113044}
  (\bibinfo {year} {2018})}\BibitemShut {NoStop}%
\bibitem [{\citenamefont {Anderson}\ \emph {et~al.}(2002)\citenamefont
  {Anderson}, \citenamefont {Robinson}, \citenamefont {Martin},\ and\
  \citenamefont {Gallagher}}]{anderson_dephasing_2002}%
  \BibitemOpen
  \bibfield  {author} {\bibinfo {author} {\bibfnamefont {W.~R.}\ \bibnamefont
  {Anderson}}, \bibinfo {author} {\bibfnamefont {M.~P.}\ \bibnamefont
  {Robinson}}, \bibinfo {author} {\bibfnamefont {J.~D.~D.}\ \bibnamefont
  {Martin}}, \ and\ \bibinfo {author} {\bibfnamefont {T.~F.}\ \bibnamefont
  {Gallagher}},\ }\bibfield  {title} {\enquote {\bibinfo {title} {Dephasing of
  resonant energy transfer in a cold {{Rydberg}} gas},}\ }\href {\doibase
  10.1103/PhysRevA.65.063404} {\bibfield  {journal} {\bibinfo  {journal} {Phys.
  Rev. A}\ }\textbf {\bibinfo {volume} {65}},\ \bibinfo {pages} {063404}
  (\bibinfo {year} {2002})}\BibitemShut {NoStop}%
\bibitem [{\citenamefont {Ravets}\ \emph {et~al.}(2014)\citenamefont {Ravets},
  \citenamefont {Labuhn}, \citenamefont {Barredo}, \citenamefont {B{\'e}guin},
  \citenamefont {Lahaye},\ and\ \citenamefont
  {Browaeys}}]{ravets_coherent_2014}%
  \BibitemOpen
  \bibfield  {author} {\bibinfo {author} {\bibfnamefont {Sylvain}\ \bibnamefont
  {Ravets}}, \bibinfo {author} {\bibfnamefont {Henning}\ \bibnamefont
  {Labuhn}}, \bibinfo {author} {\bibfnamefont {Daniel}\ \bibnamefont
  {Barredo}}, \bibinfo {author} {\bibfnamefont {Lucas}\ \bibnamefont
  {B{\'e}guin}}, \bibinfo {author} {\bibfnamefont {Thierry}\ \bibnamefont
  {Lahaye}}, \ and\ \bibinfo {author} {\bibfnamefont {Antoine}\ \bibnamefont
  {Browaeys}},\ }\bibfield  {title} {\enquote {\bibinfo {title} {Coherent
  dipole-dipole coupling between two single {{Rydberg}} atoms at an
  electrically-tuned {{F{\"o}rster}} resonance},}\ }\href {\doibase
  10.1038/nphys3119} {\bibfield  {journal} {\bibinfo  {journal} {Nat Phys}\
  }\textbf {\bibinfo {volume} {10}},\ \bibinfo {pages} {914--917} (\bibinfo
  {year} {2014})}\BibitemShut {NoStop}%
\bibitem [{\citenamefont {Ravets}\ \emph {et~al.}(2015)\citenamefont {Ravets},
  \citenamefont {Labuhn}, \citenamefont {Barredo}, \citenamefont {Lahaye},\
  and\ \citenamefont {Browaeys}}]{ravets_measurement_2015}%
  \BibitemOpen
  \bibfield  {author} {\bibinfo {author} {\bibfnamefont {Sylvain}\ \bibnamefont
  {Ravets}}, \bibinfo {author} {\bibfnamefont {Henning}\ \bibnamefont
  {Labuhn}}, \bibinfo {author} {\bibfnamefont {Daniel}\ \bibnamefont
  {Barredo}}, \bibinfo {author} {\bibfnamefont {Thierry}\ \bibnamefont
  {Lahaye}}, \ and\ \bibinfo {author} {\bibfnamefont {Antoine}\ \bibnamefont
  {Browaeys}},\ }\bibfield  {title} {\enquote {\bibinfo {title} {Measurement of
  the angular dependence of the dipole-dipole interaction between two
  individual {{Rydberg}} atoms at a {{F{\"o}rster}} resonance},}\ }\href
  {\doibase 10.1103/PhysRevA.92.020701} {\bibfield  {journal} {\bibinfo
  {journal} {Phys. Rev. A}\ }\textbf {\bibinfo {volume} {92}},\ \bibinfo
  {pages} {020701(R)} (\bibinfo {year} {2015})}\BibitemShut {NoStop}%
\bibitem [{\citenamefont {{van Ditzhuijzen}}\ \emph {et~al.}(2008)\citenamefont
  {{van Ditzhuijzen}}, \citenamefont {Koenderink}, \citenamefont
  {Hern{\'a}ndez}, \citenamefont {Robicheaux}, \citenamefont {Noordam},\ and\
  \citenamefont {{van Linden van den
  Heuvell}}}]{van_ditzhuijzen_spatially_2008}%
  \BibitemOpen
  \bibfield  {author} {\bibinfo {author} {\bibfnamefont {C.~S.~E.}\
  \bibnamefont {{van Ditzhuijzen}}}, \bibinfo {author} {\bibfnamefont {A.~F.}\
  \bibnamefont {Koenderink}}, \bibinfo {author} {\bibfnamefont {J.~V.}\
  \bibnamefont {Hern{\'a}ndez}}, \bibinfo {author} {\bibfnamefont
  {F.}~\bibnamefont {Robicheaux}}, \bibinfo {author} {\bibfnamefont {L.~D.}\
  \bibnamefont {Noordam}}, \ and\ \bibinfo {author} {\bibfnamefont {H.B.}\
  \bibnamefont {{van Linden van den Heuvell}}},\ }\bibfield  {title} {\enquote
  {\bibinfo {title} {Spatially {{Resolved Observation}} of {{Dipole}}-{{Dipole
  Interaction}} between {{Rydberg Atoms}}},}\ }\href {\doibase
  10.1103/PhysRevLett.100.243201} {\bibfield  {journal} {\bibinfo  {journal}
  {Phys. Rev. Lett.}\ }\textbf {\bibinfo {volume} {100}},\ \bibinfo {pages}
  {243201} (\bibinfo {year} {2008})}\BibitemShut {NoStop}%
\bibitem [{\citenamefont {Fahey}\ \emph {et~al.}(2015)\citenamefont {Fahey},
  \citenamefont {Carroll},\ and\ \citenamefont {Noel}}]{fahey_imaging_2015}%
  \BibitemOpen
  \bibfield  {author} {\bibinfo {author} {\bibfnamefont {Donald~P.}\
  \bibnamefont {Fahey}}, \bibinfo {author} {\bibfnamefont {Thomas~J.}\
  \bibnamefont {Carroll}}, \ and\ \bibinfo {author} {\bibfnamefont
  {Michael~W.}\ \bibnamefont {Noel}},\ }\bibfield  {title} {\enquote {\bibinfo
  {title} {Imaging the dipole-dipole energy exchange between ultracold rubidium
  {{Rydberg}} atoms},}\ }\href {\doibase 10.1103/PhysRevA.91.062702} {\bibfield
   {journal} {\bibinfo  {journal} {Phys. Rev. A}\ }\textbf {\bibinfo {volume}
  {91}},\ \bibinfo {pages} {062702} (\bibinfo {year} {2015})}\BibitemShut
  {NoStop}%
\bibitem [{\citenamefont {Bigelow}\ \emph {et~al.}(2016)\citenamefont
  {Bigelow}, \citenamefont {Paul}, \citenamefont {Peleg}, \citenamefont
  {Sanford}, \citenamefont {Carroll},\ and\ \citenamefont
  {Noel}}]{bigelow_simulations_2016}%
  \BibitemOpen
  \bibfield  {author} {\bibinfo {author} {\bibfnamefont {Jacob~L.}\
  \bibnamefont {Bigelow}}, \bibinfo {author} {\bibfnamefont {Jacob~T.}\
  \bibnamefont {Paul}}, \bibinfo {author} {\bibfnamefont {Matan}\ \bibnamefont
  {Peleg}}, \bibinfo {author} {\bibfnamefont {Veronica~L.}\ \bibnamefont
  {Sanford}}, \bibinfo {author} {\bibfnamefont {Thomas~J.}\ \bibnamefont
  {Carroll}}, \ and\ \bibinfo {author} {\bibfnamefont {Michael~W.}\
  \bibnamefont {Noel}},\ }\bibfield  {title} {\enquote {\bibinfo {title}
  {Simulations of the angular dependence of the dipole\textendash{}dipole
  interaction among {{Rydberg}} atoms},}\ }\href {\doibase
  10.1088/0953-4075/49/16/164003} {\bibfield  {journal} {\bibinfo  {journal}
  {J. Phys. B: At. Mol. Opt. Phys.}\ }\textbf {\bibinfo {volume} {49}},\
  \bibinfo {pages} {164003} (\bibinfo {year} {2016})}\BibitemShut {NoStop}%
\bibitem [{\citenamefont {Robicheaux}\ and\ \citenamefont
  {Gill}(2014)}]{robicheaux_effect_2014}%
  \BibitemOpen
  \bibfield  {author} {\bibinfo {author} {\bibfnamefont {F.}~\bibnamefont
  {Robicheaux}}\ and\ \bibinfo {author} {\bibfnamefont {N.~M.}\ \bibnamefont
  {Gill}},\ }\bibfield  {title} {\enquote {\bibinfo {title} {Effect of random
  positions for coherent dipole transport},}\ }\href {\doibase
  10.1103/PhysRevA.89.053429} {\bibfield  {journal} {\bibinfo  {journal} {Phys.
  Rev. A}\ }\textbf {\bibinfo {volume} {89}},\ \bibinfo {pages} {053429}
  (\bibinfo {year} {2014})}\BibitemShut {NoStop}%
\bibitem [{\citenamefont {Orioli}\ \emph {et~al.}(2018)\citenamefont {Orioli},
  \citenamefont {Signoles}, \citenamefont {Wildhagen}, \citenamefont
  {G{\"u}nter}, \citenamefont {Berges}, \citenamefont {Whitlock},\ and\
  \citenamefont {Weidem{\"u}ller}}]{orioli_relaxation_2018}%
  \BibitemOpen
  \bibfield  {author} {\bibinfo {author} {\bibfnamefont {A.~P.}\ \bibnamefont
  {Orioli}}, \bibinfo {author} {\bibfnamefont {A.}~\bibnamefont {Signoles}},
  \bibinfo {author} {\bibfnamefont {H.}~\bibnamefont {Wildhagen}}, \bibinfo
  {author} {\bibfnamefont {G.}~\bibnamefont {G{\"u}nter}}, \bibinfo {author}
  {\bibfnamefont {J.}~\bibnamefont {Berges}}, \bibinfo {author} {\bibfnamefont
  {S.}~\bibnamefont {Whitlock}}, \ and\ \bibinfo {author} {\bibfnamefont
  {M.}~\bibnamefont {Weidem{\"u}ller}},\ }\bibfield  {title} {\enquote
  {\bibinfo {title} {Relaxation of an {{Isolated Dipolar}}-{{Interacting
  Rydberg Quantum Spin System}}},}\ }\href {\doibase
  10.1103/PhysRevLett.120.063601} {\bibfield  {journal} {\bibinfo  {journal}
  {Phys. Rev. Lett.}\ }\textbf {\bibinfo {volume} {120}},\ \bibinfo {pages}
  {063601} (\bibinfo {year} {2018})}\BibitemShut {NoStop}%
\bibitem [{\citenamefont {Kucsko}\ \emph {et~al.}(2018)\citenamefont {Kucsko},
  \citenamefont {Choi}, \citenamefont {Choi}, \citenamefont {Maurer},
  \citenamefont {Zhou}, \citenamefont {Landig}, \citenamefont {Sumiya},
  \citenamefont {Onoda}, \citenamefont {Isoya}, \citenamefont {Jelezko},
  \citenamefont {Demler}, \citenamefont {Yao},\ and\ \citenamefont
  {Lukin}}]{kucsko_critical_2018}%
  \BibitemOpen
  \bibfield  {author} {\bibinfo {author} {\bibfnamefont {G.}~\bibnamefont
  {Kucsko}}, \bibinfo {author} {\bibfnamefont {S.}~\bibnamefont {Choi}},
  \bibinfo {author} {\bibfnamefont {J.}~\bibnamefont {Choi}}, \bibinfo {author}
  {\bibfnamefont {P.~C.}\ \bibnamefont {Maurer}}, \bibinfo {author}
  {\bibfnamefont {H.}~\bibnamefont {Zhou}}, \bibinfo {author} {\bibfnamefont
  {R.}~\bibnamefont {Landig}}, \bibinfo {author} {\bibfnamefont
  {H.}~\bibnamefont {Sumiya}}, \bibinfo {author} {\bibfnamefont
  {S.}~\bibnamefont {Onoda}}, \bibinfo {author} {\bibfnamefont
  {J.}~\bibnamefont {Isoya}}, \bibinfo {author} {\bibfnamefont
  {F.}~\bibnamefont {Jelezko}}, \bibinfo {author} {\bibfnamefont
  {E.}~\bibnamefont {Demler}}, \bibinfo {author} {\bibfnamefont {N.~Y.}\
  \bibnamefont {Yao}}, \ and\ \bibinfo {author} {\bibfnamefont {M.~D.}\
  \bibnamefont {Lukin}},\ }\bibfield  {title} {\enquote {\bibinfo {title}
  {Critical {{Thermalization}} of a {{Disordered Dipolar Spin System}} in
  {{Diamond}}},}\ }\href {\doibase 10.1103/PhysRevLett.121.023601} {\bibfield
  {journal} {\bibinfo  {journal} {Phys. Rev. Lett.}\ }\textbf {\bibinfo
  {volume} {121}},\ \bibinfo {pages} {023601} (\bibinfo {year}
  {2018})}\BibitemShut {NoStop}%
\bibitem [{\citenamefont {Xu}\ \emph {et~al.}(2018)\citenamefont {Xu},
  \citenamefont {Chen}, \citenamefont {Zeng}, \citenamefont {Zhang},
  \citenamefont {Song}, \citenamefont {Liu}, \citenamefont {Guo}, \citenamefont
  {Zhang}, \citenamefont {Xu}, \citenamefont {Deng}, \citenamefont {Huang},
  \citenamefont {Wang}, \citenamefont {Zhu}, \citenamefont {Zheng},\ and\
  \citenamefont {Fan}}]{xu_emulating_2018}%
  \BibitemOpen
  \bibfield  {author} {\bibinfo {author} {\bibfnamefont {Kai}\ \bibnamefont
  {Xu}}, \bibinfo {author} {\bibfnamefont {Jin-Jun}\ \bibnamefont {Chen}},
  \bibinfo {author} {\bibfnamefont {Yu}~\bibnamefont {Zeng}}, \bibinfo {author}
  {\bibfnamefont {Yu-Ran}\ \bibnamefont {Zhang}}, \bibinfo {author}
  {\bibfnamefont {Chao}\ \bibnamefont {Song}}, \bibinfo {author} {\bibfnamefont
  {Wuxin}\ \bibnamefont {Liu}}, \bibinfo {author} {\bibfnamefont {Qiujiang}\
  \bibnamefont {Guo}}, \bibinfo {author} {\bibfnamefont {Pengfei}\ \bibnamefont
  {Zhang}}, \bibinfo {author} {\bibfnamefont {Da}~\bibnamefont {Xu}}, \bibinfo
  {author} {\bibfnamefont {Hui}\ \bibnamefont {Deng}}, \bibinfo {author}
  {\bibfnamefont {Keqiang}\ \bibnamefont {Huang}}, \bibinfo {author}
  {\bibfnamefont {H.}~\bibnamefont {Wang}}, \bibinfo {author} {\bibfnamefont
  {Xiaobo}\ \bibnamefont {Zhu}}, \bibinfo {author} {\bibfnamefont {Dongning}\
  \bibnamefont {Zheng}}, \ and\ \bibinfo {author} {\bibfnamefont {Heng}\
  \bibnamefont {Fan}},\ }\bibfield  {title} {\enquote {\bibinfo {title}
  {Emulating {{Many}}-{{Body Localization}} with a {{Superconducting Quantum
  Processor}}},}\ }\href {\doibase 10.1103/PhysRevLett.120.050507} {\bibfield
  {journal} {\bibinfo  {journal} {Phys. Rev. Lett.}\ }\textbf {\bibinfo
  {volume} {120}},\ \bibinfo {pages} {050507} (\bibinfo {year}
  {2018})}\BibitemShut {NoStop}%
\bibitem [{\citenamefont {Smith}\ \emph {et~al.}(2016)\citenamefont {Smith},
  \citenamefont {Lee}, \citenamefont {Richerme}, \citenamefont {Neyenhuis},
  \citenamefont {Hess}, \citenamefont {Hauke}, \citenamefont {Heyl},
  \citenamefont {Huse},\ and\ \citenamefont {Monroe}}]{smith_many-body_2016}%
  \BibitemOpen
  \bibfield  {author} {\bibinfo {author} {\bibfnamefont {J.}~\bibnamefont
  {Smith}}, \bibinfo {author} {\bibfnamefont {A.}~\bibnamefont {Lee}}, \bibinfo
  {author} {\bibfnamefont {P.}~\bibnamefont {Richerme}}, \bibinfo {author}
  {\bibfnamefont {B.}~\bibnamefont {Neyenhuis}}, \bibinfo {author}
  {\bibfnamefont {P.~W.}\ \bibnamefont {Hess}}, \bibinfo {author}
  {\bibfnamefont {P.}~\bibnamefont {Hauke}}, \bibinfo {author} {\bibfnamefont
  {M.}~\bibnamefont {Heyl}}, \bibinfo {author} {\bibfnamefont {D.~A.}\
  \bibnamefont {Huse}}, \ and\ \bibinfo {author} {\bibfnamefont
  {C.}~\bibnamefont {Monroe}},\ }\bibfield  {title} {\enquote {\bibinfo {title}
  {Many-body localization in a quantum simulator with programmable random
  disorder},}\ }\href {\doibase 10.1038/nphys3783} {\bibfield  {journal}
  {\bibinfo  {journal} {Nature Physics}\ }\textbf {\bibinfo {volume} {12}},\
  \bibinfo {pages} {907--911} (\bibinfo {year} {2016})}\BibitemShut {NoStop}%
\bibitem [{\citenamefont {Gurian}\ \emph {et~al.}(2012)\citenamefont {Gurian},
  \citenamefont {Cheinet}, \citenamefont {Huillery}, \citenamefont {Fioretti},
  \citenamefont {Zhao}, \citenamefont {Gould}, \citenamefont {Comparat},\ and\
  \citenamefont {Pillet}}]{gurian_observation_2012}%
  \BibitemOpen
  \bibfield  {author} {\bibinfo {author} {\bibfnamefont {J.~H.}\ \bibnamefont
  {Gurian}}, \bibinfo {author} {\bibfnamefont {P.}~\bibnamefont {Cheinet}},
  \bibinfo {author} {\bibfnamefont {P.}~\bibnamefont {Huillery}}, \bibinfo
  {author} {\bibfnamefont {A.}~\bibnamefont {Fioretti}}, \bibinfo {author}
  {\bibfnamefont {J.}~\bibnamefont {Zhao}}, \bibinfo {author} {\bibfnamefont
  {P.~L.}\ \bibnamefont {Gould}}, \bibinfo {author} {\bibfnamefont
  {D.}~\bibnamefont {Comparat}}, \ and\ \bibinfo {author} {\bibfnamefont
  {P.}~\bibnamefont {Pillet}},\ }\bibfield  {title} {\enquote {\bibinfo {title}
  {Observation of a {{Resonant Four}}-{{Body Interaction}} in {{Cold Cesium
  Rydberg Atoms}}},}\ }\href {\doibase 10.1103/PhysRevLett.108.023005}
  {\bibfield  {journal} {\bibinfo  {journal} {Phys. Rev. Lett.}\ }\textbf
  {\bibinfo {volume} {108}},\ \bibinfo {pages} {023005} (\bibinfo {year}
  {2012})}\BibitemShut {NoStop}%
\bibitem [{\citenamefont {Faoro}\ \emph {et~al.}(2015)\citenamefont {Faoro},
  \citenamefont {Pelle}, \citenamefont {Zuliani}, \citenamefont {Cheinet},
  \citenamefont {Arimondo},\ and\ \citenamefont
  {Pillet}}]{faoro_borromean_2015}%
  \BibitemOpen
  \bibfield  {author} {\bibinfo {author} {\bibfnamefont {R.}~\bibnamefont
  {Faoro}}, \bibinfo {author} {\bibfnamefont {B.}~\bibnamefont {Pelle}},
  \bibinfo {author} {\bibfnamefont {A.}~\bibnamefont {Zuliani}}, \bibinfo
  {author} {\bibfnamefont {P.}~\bibnamefont {Cheinet}}, \bibinfo {author}
  {\bibfnamefont {E.}~\bibnamefont {Arimondo}}, \ and\ \bibinfo {author}
  {\bibfnamefont {P.}~\bibnamefont {Pillet}},\ }\bibfield  {title} {\enquote
  {\bibinfo {title} {Borromean three-body {{FRET}} in frozen {{Rydberg}}
  gases},}\ }\href {\doibase 10.1038/ncomms9173} {\bibfield  {journal}
  {\bibinfo  {journal} {Nature Communications}\ }\textbf {\bibinfo {volume}
  {6}},\ \bibinfo {pages} {1--7} (\bibinfo {year} {2015})}\BibitemShut
  {NoStop}%
\bibitem [{\citenamefont {Tretyakov}\ \emph {et~al.}(2017)\citenamefont
  {Tretyakov}, \citenamefont {Beterov}, \citenamefont {Yakshina}, \citenamefont
  {Entin}, \citenamefont {Ryabtsev}, \citenamefont {Cheinet},\ and\
  \citenamefont {Pillet}}]{tretyakov_observation_2017}%
  \BibitemOpen
  \bibfield  {author} {\bibinfo {author} {\bibfnamefont {D.~B.}\ \bibnamefont
  {Tretyakov}}, \bibinfo {author} {\bibfnamefont {I.~I.}\ \bibnamefont
  {Beterov}}, \bibinfo {author} {\bibfnamefont {E.~A.}\ \bibnamefont
  {Yakshina}}, \bibinfo {author} {\bibfnamefont {V.~M.}\ \bibnamefont {Entin}},
  \bibinfo {author} {\bibfnamefont {I.~I.}\ \bibnamefont {Ryabtsev}}, \bibinfo
  {author} {\bibfnamefont {P.}~\bibnamefont {Cheinet}}, \ and\ \bibinfo
  {author} {\bibfnamefont {P.}~\bibnamefont {Pillet}},\ }\bibfield  {title}
  {\enquote {\bibinfo {title} {Observation of the {{Borromean Three}}-{{Body
  F{\"o}rster Resonances}} for {{Three Interacting Rb Rydberg Atoms}}},}\
  }\href {\doibase 10.1103/PhysRevLett.119.173402} {\bibfield  {journal}
  {\bibinfo  {journal} {Phys. Rev. Lett.}\ }\textbf {\bibinfo {volume} {119}},\
  \bibinfo {pages} {173402} (\bibinfo {year} {2017})}\BibitemShut {NoStop}%
\bibitem [{\citenamefont {Beterov}\ \emph {et~al.}(2018)\citenamefont
  {Beterov}, \citenamefont {Ashkarin}, \citenamefont {Yakshina}, \citenamefont
  {Tretyakov}, \citenamefont {Entin}, \citenamefont {Ryabtsev}, \citenamefont
  {Cheinet}, \citenamefont {Pillet},\ and\ \citenamefont
  {Saffman}}]{beterov_fast_2018}%
  \BibitemOpen
  \bibfield  {author} {\bibinfo {author} {\bibfnamefont {I.~I.}\ \bibnamefont
  {Beterov}}, \bibinfo {author} {\bibfnamefont {I.~N.}\ \bibnamefont
  {Ashkarin}}, \bibinfo {author} {\bibfnamefont {E.~A.}\ \bibnamefont
  {Yakshina}}, \bibinfo {author} {\bibfnamefont {D.~B.}\ \bibnamefont
  {Tretyakov}}, \bibinfo {author} {\bibfnamefont {V.~M.}\ \bibnamefont
  {Entin}}, \bibinfo {author} {\bibfnamefont {I.~I.}\ \bibnamefont {Ryabtsev}},
  \bibinfo {author} {\bibfnamefont {P.}~\bibnamefont {Cheinet}}, \bibinfo
  {author} {\bibfnamefont {P.}~\bibnamefont {Pillet}}, \ and\ \bibinfo {author}
  {\bibfnamefont {M.}~\bibnamefont {Saffman}},\ }\bibfield  {title} {\enquote
  {\bibinfo {title} {Fast three-qubit {{Toffoli}} quantum gate based on
  three-body {{F{\"o}rster}} resonances in {{Rydberg}} atoms},}\ }\href
  {\doibase 10.1103/PhysRevA.98.042704} {\bibfield  {journal} {\bibinfo
  {journal} {Phys. Rev. A}\ }\textbf {\bibinfo {volume} {98}},\ \bibinfo
  {pages} {042704} (\bibinfo {year} {2018})}\BibitemShut {NoStop}%
\bibitem [{\citenamefont {Ryabtsev}\ \emph {et~al.}(2018)\citenamefont
  {Ryabtsev}, \citenamefont {Beterov}, \citenamefont {Tretyakov}, \citenamefont
  {Yakshina}, \citenamefont {Entin}, \citenamefont {Cheinet},\ and\
  \citenamefont {Pillet}}]{ryabtsev_coherence_2018}%
  \BibitemOpen
  \bibfield  {author} {\bibinfo {author} {\bibfnamefont {I.~I.}\ \bibnamefont
  {Ryabtsev}}, \bibinfo {author} {\bibfnamefont {I.~I.}\ \bibnamefont
  {Beterov}}, \bibinfo {author} {\bibfnamefont {D.~B.}\ \bibnamefont
  {Tretyakov}}, \bibinfo {author} {\bibfnamefont {E.~A.}\ \bibnamefont
  {Yakshina}}, \bibinfo {author} {\bibfnamefont {V.~M.}\ \bibnamefont {Entin}},
  \bibinfo {author} {\bibfnamefont {P.}~\bibnamefont {Cheinet}}, \ and\
  \bibinfo {author} {\bibfnamefont {P.}~\bibnamefont {Pillet}},\ }\bibfield
  {title} {\enquote {\bibinfo {title} {Coherence of three-body {{F{\"o}rster}}
  resonances in {{Rydberg}} atoms},}\ }\href {\doibase
  10.1103/PhysRevA.98.052703} {\bibfield  {journal} {\bibinfo  {journal} {Phys.
  Rev. A}\ }\textbf {\bibinfo {volume} {98}},\ \bibinfo {pages} {052703}
  (\bibinfo {year} {2018})}\BibitemShut {NoStop}%
\bibitem [{\citenamefont {Gregoric}\ \emph {et~al.}(2017)\citenamefont
  {Gregoric}, \citenamefont {Kang}, \citenamefont {Liu}, \citenamefont
  {Rowley}, \citenamefont {Carroll},\ and\ \citenamefont
  {Noel}}]{gregoric_quantum_2017}%
  \BibitemOpen
  \bibfield  {author} {\bibinfo {author} {\bibfnamefont {Vincent~C.}\
  \bibnamefont {Gregoric}}, \bibinfo {author} {\bibfnamefont {Xinyue}\
  \bibnamefont {Kang}}, \bibinfo {author} {\bibfnamefont {Zhimin~Cheryl}\
  \bibnamefont {Liu}}, \bibinfo {author} {\bibfnamefont {Zoe~A.}\ \bibnamefont
  {Rowley}}, \bibinfo {author} {\bibfnamefont {Thomas~J.}\ \bibnamefont
  {Carroll}}, \ and\ \bibinfo {author} {\bibfnamefont {Michael~W.}\
  \bibnamefont {Noel}},\ }\bibfield  {title} {\enquote {\bibinfo {title}
  {Quantum control via a genetic algorithm of the field ionization pathway of a
  {{Rydberg}} electron},}\ }\href {\doibase 10.1103/PhysRevA.96.023403}
  {\bibfield  {journal} {\bibinfo  {journal} {Phys. Rev. A}\ }\textbf {\bibinfo
  {volume} {96}},\ \bibinfo {pages} {023403} (\bibinfo {year}
  {2017})}\BibitemShut {NoStop}%
\bibitem [{\citenamefont {Gregoric}\ \emph {et~al.}(2018)\citenamefont
  {Gregoric}, \citenamefont {Bennett}, \citenamefont {Gualtieri}, \citenamefont
  {Kannad}, \citenamefont {Liu}, \citenamefont {Rowley}, \citenamefont
  {Carroll},\ and\ \citenamefont {Noel}}]{gregoric_improving_2018}%
  \BibitemOpen
  \bibfield  {author} {\bibinfo {author} {\bibfnamefont {Vincent~C.}\
  \bibnamefont {Gregoric}}, \bibinfo {author} {\bibfnamefont {Jason~J.}\
  \bibnamefont {Bennett}}, \bibinfo {author} {\bibfnamefont {Bianca~R.}\
  \bibnamefont {Gualtieri}}, \bibinfo {author} {\bibfnamefont {Ankitha}\
  \bibnamefont {Kannad}}, \bibinfo {author} {\bibfnamefont {Zhimin~Cheryl}\
  \bibnamefont {Liu}}, \bibinfo {author} {\bibfnamefont {Zoe~A.}\ \bibnamefont
  {Rowley}}, \bibinfo {author} {\bibfnamefont {Thomas~J.}\ \bibnamefont
  {Carroll}}, \ and\ \bibinfo {author} {\bibfnamefont {Michael~W.}\
  \bibnamefont {Noel}},\ }\bibfield  {title} {\enquote {\bibinfo {title}
  {Improving the state selectivity of field ionization with quantum control},}\
  }\href {\doibase 10.1103/PhysRevA.98.063404} {\bibfield  {journal} {\bibinfo
  {journal} {Phys. Rev. A}\ }\textbf {\bibinfo {volume} {98}},\ \bibinfo
  {pages} {063404} (\bibinfo {year} {2018})}\BibitemShut {NoStop}%
\bibitem [{\citenamefont {Gregoric}\ \emph {et~al.}(2019)\citenamefont
  {Gregoric}, \citenamefont {Bennett}, \citenamefont {Gualtieri}, \citenamefont
  {Hastings}, \citenamefont {Kannad}, \citenamefont {Liu}, \citenamefont
  {Rabinowitz}, \citenamefont {Rowley}, \citenamefont {Wang}, \citenamefont
  {Yoast}, \citenamefont {Carroll},\ and\ \citenamefont
  {Noel}}]{gregoric_perturbed_2019}%
  \BibitemOpen
  \bibfield  {author} {\bibinfo {author} {\bibfnamefont {Vincent~C.}\
  \bibnamefont {Gregoric}}, \bibinfo {author} {\bibfnamefont {Jason~J.}\
  \bibnamefont {Bennett}}, \bibinfo {author} {\bibfnamefont {Bianca~R.}\
  \bibnamefont {Gualtieri}}, \bibinfo {author} {\bibfnamefont {Hannah~P.}\
  \bibnamefont {Hastings}}, \bibinfo {author} {\bibfnamefont {Ankitha}\
  \bibnamefont {Kannad}}, \bibinfo {author} {\bibfnamefont {Zhimin~Cheryl}\
  \bibnamefont {Liu}}, \bibinfo {author} {\bibfnamefont {Maia~R.}\ \bibnamefont
  {Rabinowitz}}, \bibinfo {author} {\bibfnamefont {Zoe~A.}\ \bibnamefont
  {Rowley}}, \bibinfo {author} {\bibfnamefont {Miao}\ \bibnamefont {Wang}},
  \bibinfo {author} {\bibfnamefont {Lauren}\ \bibnamefont {Yoast}}, \bibinfo
  {author} {\bibfnamefont {Thomas~J.}\ \bibnamefont {Carroll}}, \ and\ \bibinfo
  {author} {\bibfnamefont {Michael~W.}\ \bibnamefont {Noel}},\ }\bibfield
  {title} {\enquote {\bibinfo {title} {Perturbed {{Field Ionization}} for
  {{Improved State Selectivity}}},}\ }\href {\doibase 10.1088/1361-6455/ab707a}
  {\bibfield  {journal} {\bibinfo  {journal} {arXiv:1908.09052 [physics,
  physics:quant-ph]}\ } (\bibinfo {year} {2019}),\
  10.1088/1361-6455/ab707a}\BibitemShut {NoStop}%
\bibitem [{\citenamefont {Edgal}\ and\ \citenamefont
  {Wiley}(1983)}]{edgal_near-neighbor_1983}%
  \BibitemOpen
  \bibfield  {author} {\bibinfo {author} {\bibfnamefont {Uduzei~F.}\
  \bibnamefont {Edgal}}\ and\ \bibinfo {author} {\bibfnamefont {J.~D.}\
  \bibnamefont {Wiley}},\ }\bibfield  {title} {\enquote {\bibinfo {title}
  {Near-neighbor configuration and impurity-cluster size distribution in a
  {{Poisson}} ensemble of monovalent impurity atoms in semiconductors},}\
  }\href {\doibase 10.1103/PhysRevB.27.4997} {\bibfield  {journal} {\bibinfo
  {journal} {Phys. Rev. B}\ }\textbf {\bibinfo {volume} {27}},\ \bibinfo
  {pages} {4997--5006} (\bibinfo {year} {1983})}\BibitemShut {NoStop}%
\bibitem [{\citenamefont
  {Chandrasekhar}(1943)}]{chandrasekhar_stochastic_1943}%
  \BibitemOpen
  \bibfield  {author} {\bibinfo {author} {\bibfnamefont {S.}~\bibnamefont
  {Chandrasekhar}},\ }\bibfield  {title} {\enquote {\bibinfo {title}
  {Stochastic {{Problems}} in {{Physics}} and {{Astronomy}}},}\ }\href
  {\doibase 10.1103/RevModPhys.15.1} {\bibfield  {journal} {\bibinfo  {journal}
  {Rev. Mod. Phys.}\ }\textbf {\bibinfo {volume} {15}},\ \bibinfo {pages}
  {1--89} (\bibinfo {year} {1943})}\BibitemShut {NoStop}%
\bibitem [{\citenamefont {Younge}\ \emph {et~al.}(2009)\citenamefont {Younge},
  \citenamefont {Reinhard}, \citenamefont {Pohl}, \citenamefont {Berman},\ and\
  \citenamefont {Raithel}}]{younge_mesoscopic_2009}%
  \BibitemOpen
  \bibfield  {author} {\bibinfo {author} {\bibfnamefont {K.~C.}\ \bibnamefont
  {Younge}}, \bibinfo {author} {\bibfnamefont {A.}~\bibnamefont {Reinhard}},
  \bibinfo {author} {\bibfnamefont {T.}~\bibnamefont {Pohl}}, \bibinfo {author}
  {\bibfnamefont {P.~R.}\ \bibnamefont {Berman}}, \ and\ \bibinfo {author}
  {\bibfnamefont {G.}~\bibnamefont {Raithel}},\ }\bibfield  {title} {\enquote
  {\bibinfo {title} {Mesoscopic {{Rydberg}} ensembles: {{Beyond}} the
  pairwise-interaction approximation},}\ }\href {\doibase
  10.1103/PhysRevA.79.043420} {\bibfield  {journal} {\bibinfo  {journal} {Phys.
  Rev. A}\ }\textbf {\bibinfo {volume} {79}},\ \bibinfo {pages} {043420}
  (\bibinfo {year} {2009})}\BibitemShut {NoStop}%
\bibitem [{\citenamefont {Cournol}\ \emph {et~al.}(2018)\citenamefont
  {Cournol}, \citenamefont {Robert}, \citenamefont {Pillet},\ and\
  \citenamefont {Vanhaecke}}]{cournol_accurate_2018}%
  \BibitemOpen
  \bibfield  {author} {\bibinfo {author} {\bibfnamefont {Anne}\ \bibnamefont
  {Cournol}}, \bibinfo {author} {\bibfnamefont {Jacques}\ \bibnamefont
  {Robert}}, \bibinfo {author} {\bibfnamefont {Pierre}\ \bibnamefont {Pillet}},
  \ and\ \bibinfo {author} {\bibfnamefont {Nicolas}\ \bibnamefont
  {Vanhaecke}},\ }\bibfield  {title} {\enquote {\bibinfo {title} {Accurate
  density measurement of a cold {{Rydberg}} gas via non collisional two-body
  transitions},}\ }\href {\doibase 10.1088/1367-2630/aad214} {\bibfield
  {journal} {\bibinfo  {journal} {New J. Phys.}\ }\textbf {\bibinfo {volume}
  {20}},\ \bibinfo {pages} {073042} (\bibinfo {year} {2018})}\BibitemShut
  {NoStop}%
\bibitem [{\citenamefont {Barredo}\ \emph {et~al.}(2015)\citenamefont
  {Barredo}, \citenamefont {Labuhn}, \citenamefont {Ravets}, \citenamefont
  {Lahaye}, \citenamefont {Browaeys},\ and\ \citenamefont
  {Adams}}]{barredo_coherent_2015}%
  \BibitemOpen
  \bibfield  {author} {\bibinfo {author} {\bibfnamefont {Daniel}\ \bibnamefont
  {Barredo}}, \bibinfo {author} {\bibfnamefont {Henning}\ \bibnamefont
  {Labuhn}}, \bibinfo {author} {\bibfnamefont {Sylvain}\ \bibnamefont
  {Ravets}}, \bibinfo {author} {\bibfnamefont {Thierry}\ \bibnamefont
  {Lahaye}}, \bibinfo {author} {\bibfnamefont {Antoine}\ \bibnamefont
  {Browaeys}}, \ and\ \bibinfo {author} {\bibfnamefont {Charles~S.}\
  \bibnamefont {Adams}},\ }\bibfield  {title} {\enquote {\bibinfo {title}
  {Coherent {{Excitation Transfer}} in a {{Spin Chain}} of {{Three Rydberg
  Atoms}}},}\ }\href {\doibase 10.1103/PhysRevLett.114.113002} {\bibfield
  {journal} {\bibinfo  {journal} {Phys. Rev. Lett.}\ }\textbf {\bibinfo
  {volume} {114}},\ \bibinfo {pages} {113002} (\bibinfo {year}
  {2015})}\BibitemShut {NoStop}%
\bibitem [{\citenamefont {Carroll}\ \emph {et~al.}(2004)\citenamefont
  {Carroll}, \citenamefont {Claringbould}, \citenamefont {Goodsell},
  \citenamefont {Lim},\ and\ \citenamefont {Noel}}]{carroll_angular_2004}%
  \BibitemOpen
  \bibfield  {author} {\bibinfo {author} {\bibfnamefont {Thomas~J.}\
  \bibnamefont {Carroll}}, \bibinfo {author} {\bibfnamefont {Katharine}\
  \bibnamefont {Claringbould}}, \bibinfo {author} {\bibfnamefont {Anne}\
  \bibnamefont {Goodsell}}, \bibinfo {author} {\bibfnamefont {M.~J.}\
  \bibnamefont {Lim}}, \ and\ \bibinfo {author} {\bibfnamefont {Michael~W.}\
  \bibnamefont {Noel}},\ }\bibfield  {title} {\enquote {\bibinfo {title}
  {Angular {{Dependence}} of the {{Dipole}}-{{Dipole Interaction}} in a
  {{Nearly One}}-{{Dimensional Sample}} of {{Rydberg Atoms}}},}\ }\href
  {\doibase 10.1103/PhysRevLett.93.153001} {\bibfield  {journal} {\bibinfo
  {journal} {Phys. Rev. Lett.}\ }\textbf {\bibinfo {volume} {93}},\ \bibinfo
  {pages} {153001} (\bibinfo {year} {2004})}\BibitemShut {NoStop}%
\bibitem [{\citenamefont {Polkovnikov}\ \emph {et~al.}(2011)\citenamefont
  {Polkovnikov}, \citenamefont {Sengupta}, \citenamefont {Silva},\ and\
  \citenamefont {Vengalattore}}]{polkovnikov_colloquium_2011}%
  \BibitemOpen
  \bibfield  {author} {\bibinfo {author} {\bibfnamefont {Anatoli}\ \bibnamefont
  {Polkovnikov}}, \bibinfo {author} {\bibfnamefont {Krishnendu}\ \bibnamefont
  {Sengupta}}, \bibinfo {author} {\bibfnamefont {Alessandro}\ \bibnamefont
  {Silva}}, \ and\ \bibinfo {author} {\bibfnamefont {Mukund}\ \bibnamefont
  {Vengalattore}},\ }\bibfield  {title} {\enquote {\bibinfo {title}
  {Colloquium: {{Nonequilibrium}} dynamics of closed interacting quantum
  systems},}\ }\href {\doibase 10.1103/RevModPhys.83.863} {\bibfield  {journal}
  {\bibinfo  {journal} {Rev. Mod. Phys.}\ }\textbf {\bibinfo {volume} {83}},\
  \bibinfo {pages} {863--883} (\bibinfo {year} {2011})}\BibitemShut {NoStop}%
\bibitem [{\citenamefont {Abanin}\ and\ \citenamefont
  {Papi{\'c}}(2017)}]{abanin_recent_2017}%
  \BibitemOpen
  \bibfield  {author} {\bibinfo {author} {\bibfnamefont {Dmitry~A.}\
  \bibnamefont {Abanin}}\ and\ \bibinfo {author} {\bibfnamefont {Zlatko}\
  \bibnamefont {Papi{\'c}}},\ }\bibfield  {title} {\enquote {\bibinfo {title}
  {Recent progress in many-body localization},}\ }\href {\doibase
  10.1002/andp.201700169} {\bibfield  {journal} {\bibinfo  {journal} {Annalen
  der Physik}\ }\textbf {\bibinfo {volume} {529}},\ \bibinfo {pages} {1700169}
  (\bibinfo {year} {2017})}\BibitemShut {NoStop}%
\bibitem [{\citenamefont {Nandkishore}\ and\ \citenamefont
  {Huse}(2015)}]{nandkishore_many-body_2015}%
  \BibitemOpen
  \bibfield  {author} {\bibinfo {author} {\bibfnamefont {Rahul}\ \bibnamefont
  {Nandkishore}}\ and\ \bibinfo {author} {\bibfnamefont {David~A.}\
  \bibnamefont {Huse}},\ }\bibfield  {title} {\enquote {\bibinfo {title}
  {Many-{{Body Localization}} and {{Thermalization}} in {{Quantum Statistical
  Mechanics}}},}\ }\href {\doibase 10.1146/annurev-conmatphys-031214-014726}
  {\bibfield  {journal} {\bibinfo  {journal} {Annual Review of Condensed Matter
  Physics}\ }\textbf {\bibinfo {volume} {6}},\ \bibinfo {pages} {15--38}
  (\bibinfo {year} {2015})}\BibitemShut {NoStop}%
\bibitem [{\citenamefont {Abanin}\ \emph {et~al.}(2019)\citenamefont {Abanin},
  \citenamefont {Altman}, \citenamefont {Bloch},\ and\ \citenamefont
  {Serbyn}}]{abanin_colloquium_2019}%
  \BibitemOpen
  \bibfield  {author} {\bibinfo {author} {\bibfnamefont {Dmitry~A.}\
  \bibnamefont {Abanin}}, \bibinfo {author} {\bibfnamefont {Ehud}\ \bibnamefont
  {Altman}}, \bibinfo {author} {\bibfnamefont {Immanuel}\ \bibnamefont
  {Bloch}}, \ and\ \bibinfo {author} {\bibfnamefont {Maksym}\ \bibnamefont
  {Serbyn}},\ }\bibfield  {title} {\enquote {\bibinfo {title} {Colloquium:
  {{Many}}-body localization, thermalization, and entanglement},}\ }\href
  {\doibase 10.1103/RevModPhys.91.021001} {\bibfield  {journal} {\bibinfo
  {journal} {Rev. Mod. Phys.}\ }\textbf {\bibinfo {volume} {91}},\ \bibinfo
  {pages} {021001} (\bibinfo {year} {2019})}\BibitemShut {NoStop}%
\bibitem [{\citenamefont {Lukin}\ \emph {et~al.}(2019)\citenamefont {Lukin},
  \citenamefont {Rispoli}, \citenamefont {Schittko}, \citenamefont {Tai},
  \citenamefont {Kaufman}, \citenamefont {Choi}, \citenamefont {Khemani},
  \citenamefont {L{\'e}onard},\ and\ \citenamefont
  {Greiner}}]{lukin_probing_2019}%
  \BibitemOpen
  \bibfield  {author} {\bibinfo {author} {\bibfnamefont {Alexander}\
  \bibnamefont {Lukin}}, \bibinfo {author} {\bibfnamefont {Matthew}\
  \bibnamefont {Rispoli}}, \bibinfo {author} {\bibfnamefont {Robert}\
  \bibnamefont {Schittko}}, \bibinfo {author} {\bibfnamefont {M.~Eric}\
  \bibnamefont {Tai}}, \bibinfo {author} {\bibfnamefont {Adam~M.}\ \bibnamefont
  {Kaufman}}, \bibinfo {author} {\bibfnamefont {Soonwon}\ \bibnamefont {Choi}},
  \bibinfo {author} {\bibfnamefont {Vedika}\ \bibnamefont {Khemani}}, \bibinfo
  {author} {\bibfnamefont {Julian}\ \bibnamefont {L{\'e}onard}}, \ and\
  \bibinfo {author} {\bibfnamefont {Markus}\ \bibnamefont {Greiner}},\
  }\bibfield  {title} {\enquote {\bibinfo {title} {Probing entanglement in a
  many-body\textendash{}localized system},}\ }\href {\doibase
  10.1126/science.aau0818} {\bibfield  {journal} {\bibinfo  {journal}
  {Science}\ }\textbf {\bibinfo {volume} {364}},\ \bibinfo {pages} {256--260}
  (\bibinfo {year} {2019})}\BibitemShut {NoStop}%
\bibitem [{\citenamefont {Sous}\ and\ \citenamefont
  {Grant}(2018)}]{sous_possible_2018}%
  \BibitemOpen
  \bibfield  {author} {\bibinfo {author} {\bibfnamefont {John}\ \bibnamefont
  {Sous}}\ and\ \bibinfo {author} {\bibfnamefont {Edward}\ \bibnamefont
  {Grant}},\ }\bibfield  {title} {\enquote {\bibinfo {title} {Possible
  {{Many}}-{{Body Localization}} in a {{Long}}-{{Lived Finite}}-{{Temperature
  Ultracold Quasineutral Molecular Plasma}}},}\ }\href {\doibase
  10.1103/PhysRevLett.120.110601} {\bibfield  {journal} {\bibinfo  {journal}
  {Phys. Rev. Lett.}\ }\textbf {\bibinfo {volume} {120}},\ \bibinfo {pages}
  {110601} (\bibinfo {year} {2018})}\BibitemShut {NoStop}%
\bibitem [{\citenamefont {Sous}\ and\ \citenamefont
  {Grant}(2019)}]{sous_many-body_2019}%
  \BibitemOpen
  \bibfield  {author} {\bibinfo {author} {\bibfnamefont {John}\ \bibnamefont
  {Sous}}\ and\ \bibinfo {author} {\bibfnamefont {Edward}\ \bibnamefont
  {Grant}},\ }\bibfield  {title} {\enquote {\bibinfo {title} {Many-body physics
  with ultracold plasmas: Quenched randomness and localization},}\ }\href
  {\doibase 10.1088/1367-2630/aaf669} {\bibfield  {journal} {\bibinfo
  {journal} {New J. Phys.}\ }\textbf {\bibinfo {volume} {21}},\ \bibinfo
  {pages} {043033} (\bibinfo {year} {2019})}\BibitemShut {NoStop}%
\bibitem [{\citenamefont {Nandkishore}\ and\ \citenamefont
  {Sondhi}(2017)}]{nandkishore_many-body_2017}%
  \BibitemOpen
  \bibfield  {author} {\bibinfo {author} {\bibfnamefont {Rahul~M.}\
  \bibnamefont {Nandkishore}}\ and\ \bibinfo {author} {\bibfnamefont {S.~L.}\
  \bibnamefont {Sondhi}},\ }\bibfield  {title} {\enquote {\bibinfo {title}
  {Many-{{Body Localization}} with {{Long}}-{{Range Interactions}}},}\ }\href
  {\doibase 10.1103/PhysRevX.7.041021} {\bibfield  {journal} {\bibinfo
  {journal} {Phys. Rev. X}\ }\textbf {\bibinfo {volume} {7}},\ \bibinfo {pages}
  {041021} (\bibinfo {year} {2017})}\BibitemShut {NoStop}%
\bibitem [{\citenamefont {T{\'a}vora}\ \emph {et~al.}(2016)\citenamefont
  {T{\'a}vora}, \citenamefont {{Torres-Herrera}},\ and\ \citenamefont
  {Santos}}]{tavora_inevitable_2016}%
  \BibitemOpen
  \bibfield  {author} {\bibinfo {author} {\bibfnamefont {Marco}\ \bibnamefont
  {T{\'a}vora}}, \bibinfo {author} {\bibfnamefont {E.~J.}\ \bibnamefont
  {{Torres-Herrera}}}, \ and\ \bibinfo {author} {\bibfnamefont {Lea~F.}\
  \bibnamefont {Santos}},\ }\bibfield  {title} {\enquote {\bibinfo {title}
  {Inevitable power-law behavior of isolated many-body quantum systems and how
  it anticipates thermalization},}\ }\href {\doibase
  10.1103/PhysRevA.94.041603} {\bibfield  {journal} {\bibinfo  {journal} {Phys.
  Rev. A}\ }\textbf {\bibinfo {volume} {94}},\ \bibinfo {pages} {041603(R)}
  (\bibinfo {year} {2016})}\BibitemShut {NoStop}%
\bibitem [{\citenamefont {T{\'a}vora}\ \emph {et~al.}(2017)\citenamefont
  {T{\'a}vora}, \citenamefont {{Torres-Herrera}},\ and\ \citenamefont
  {Santos}}]{tavora_power-law_2017}%
  \BibitemOpen
  \bibfield  {author} {\bibinfo {author} {\bibfnamefont {Marco}\ \bibnamefont
  {T{\'a}vora}}, \bibinfo {author} {\bibfnamefont {E.~J.}\ \bibnamefont
  {{Torres-Herrera}}}, \ and\ \bibinfo {author} {\bibfnamefont {Lea~F.}\
  \bibnamefont {Santos}},\ }\bibfield  {title} {\enquote {\bibinfo {title}
  {Power-law decay exponents: {{A}} dynamical criterion for predicting
  thermalization},}\ }\href {\doibase 10.1103/PhysRevA.95.013604} {\bibfield
  {journal} {\bibinfo  {journal} {Phys. Rev. A}\ }\textbf {\bibinfo {volume}
  {95}},\ \bibinfo {pages} {013604} (\bibinfo {year} {2017})}\BibitemShut
  {NoStop}%
\end{thebibliography}%

\end{document}